\documentclass[conference]{IEEEtran}
\usepackage{tikz}
\usepackage{amsmath}
\usepackage{subcaption}
\usepackage{filecontents}
\usepackage{amsfonts}
\usepackage{algorithm}
\usepackage{algorithmic}
\usepackage{url}
\usepackage{colortbl}
\usepackage{multirow}
\usepackage{hyperref}
\usepackage{amssymb}
\usepackage{booktabs}
\usepackage{float}
\usepackage{pifont}
\usepackage{amsthm}
\usepackage{cite}
\theoremstyle{definition}

\newtheorem{theorem}{Theorem}

\definecolor{c2}{HTML}{F9CE9C}
\definecolor{c1}{HTML}{EDB8B0}
\definecolor{c3}{HTML}{D9B9D4}
\newcommand{\cmark}{\ding{51}}  % √
\newcommand{\xmark}{\ding{55}}  % ✗

\ifCLASSINFOpdf
\else
\fi

\begin{document}
\title{Multi-Party Private Set Intersection from Intersection-Gated Oblivious Pseudorandom Function}

\author{
\IEEEauthorblockN{
Guowei~Ling\IEEEauthorrefmark{1}\IEEEauthorrefmark{2},
Peng~Tang\IEEEauthorrefmark{1},
Jingwei~Hu\IEEEauthorrefmark{3},
Liyao~Xiang\IEEEauthorrefmark{1},
Fei~Tang\IEEEauthorrefmark{4},\\
Jinyong~Shan\IEEEauthorrefmark{5},
Weidong~Qiu\IEEEauthorrefmark{1}\IEEEauthorrefmark{6},
San~Ling\IEEEauthorrefmark{2}, and
Huaxiong~Wang\IEEEauthorrefmark{2}
}
\IEEEauthorblockA{
\IEEEauthorrefmark{1}Shanghai Jiao Tong University, China \quad
\IEEEauthorrefmark{2}Nanyang Technological University, Singapore\\
\IEEEauthorrefmark{3}Beijing Institute of Technology, China \quad
\IEEEauthorrefmark{4}Chongqing University of Posts and Telecommunications, China\\
\IEEEauthorrefmark{5}Beijing Smartchip Microelectronics Technology Co., Ltd., China\\
Emails: \{gw\_ling,tangpeng,xiangliyao08,qiuwd\}@sjtu.edu.cn,
\{davidhu,lingsan,hxwang\}@ntu.edu.sg\\
tangfei@cqupt.edu.cn, shanjinyong@sgchip.sgcc.com.cn\\
\IEEEauthorrefmark{6}Corresponding author.
}
}

\IEEEoverridecommandlockouts
\makeatletter\def\@IEEEpubidpullup{7.5\baselineskip}\makeatother
\IEEEpubid{\parbox{\columnwidth}{
\textit{Revised preprint. Accepted at NDSS 2027 (Summer cycle).}\\[2pt]
Network and Distributed System Security (NDSS) Symposium 2027\\
22--26 March 2027, Seoul, Republic of Korea\\
ISBN 978-1-970672-09-1\\
https://dx.doi.org/10.14722/ndss.2027.230484\\
www.ndss-symposium.org
}
\hspace{\columnsep}\makebox[\columnwidth]{}}
\maketitle


\begin{abstract}
Multi-party private set intersection (MPSI) enables multiple parties to compute the intersection of their private sets without revealing any information beyond the intersection. We introduce the intersection-gated oblivious pseudorandom function (IGOPRF), which allows the receiver to obtain a sender-consistent pseudorandom label only for elements belonging to the intersection of all parties' sets. This functionality enables the construction of MPSI protocols in both the semi-honest and malicious settings. We further extend IGOPRF to an intersection-gated oblivious programmable pseudorandom function to support MPSI cardinality, threshold MPSI, and MPSI-sum. Our constructions are secure under the designated sender-receiver non-collusion assumption. In the malicious setting, our protocol tolerates all corruption patterns admissible under this assumption. Online measurements show that our maliciously secure protocol achieves speedups of up to $23.0\times$ in LAN and $6.4\times$ in WAN over the state-of-the-art maliciously secure MPSI protocol, while reducing communication by up to $3.9\times$. In the semi-honest setting, our protocol achieves a speedup of up to $22.0\times$ in LAN over a state-of-the-art semi-honest MPSI protocol in settings with large inputs or high corruption tolerance. Overall, our constructions combine practical efficiency, straightforward extensibility to richer functionalities, and broader malicious-security coverage.
\end{abstract}


% \fi
\IEEEpeerreviewmaketitle

\section{Introduction}

Private set intersection (PSI) enables parties, each holding a private set of elements, to compute the intersection of their sets without revealing anything beyond the intersection itself. In recent years, two-party PSI has become highly efficient in practice, with high-performance implementations that provide strong cryptographic security \cite{kkrt16,prty19,cm20,prty20,prty21,rs21,rr22,bpsy23,tdsc25,ling26upsi,ling26psi}. Many modern two-party PSI protocols are built around the oblivious pseudorandom function (OPRF) paradigm. At a high level, an OPRF provides the main building block for constructing an efficient two-party PSI protocol.

The power of this paradigm is well demonstrated by the protocols in \cite{rs21,rr22}, which achieve remarkable efficiency, computing the intersection of sets containing millions of elements in less than one second in their reported experiments. Beyond their efficiency, these protocols support both semi-honest and malicious security and can be extended to a range of PSI-related functionalities. Such flexibility is important in practice, as parties are often interested in more than simply learning which elements they have in common. They may also want to perform further computations on these elements, such as computing the cardinality of the intersection or aggregating the payloads associated with the common elements. Taken together, these advances show how far two-party PSI has come over the past several years.

Despite several notable advances in recent years, MPSI is still less mature than its two-party counterpart. One reason is that the two areas have followed quite different paths in protocol design. For two-party PSI, OPRF provides a natural and convenient starting point: once an OPRF is available, it can be used to build an efficient PSI protocol. However, MPSI does not have an equally straightforward approach. Existing protocols \cite{kmp17,cdg21,wyc24,glw24,gga25,dcz25} typically combine several building blocks, with each construction relying on its own combination of components. As a result, it remains challenging to achieve strong security, high efficiency, and support for richer functionalities within a single MPSI construction. As shown in Table~\ref{workcomparison}, existing protocols have yet to bring these three aspects together.

Moreover, malicious security remains limited among existing efficient MPSI protocols. Only \cite{nty21,glw24} provide malicious security, and even \cite{glw24} considers malicious clients only, rather than malicious corruption of any party. Therefore, we aim to address these limitations of existing MPSI protocols. To this end, we introduce a new primitive, called the intersection-gated OPRF (IGOPRF), as a basis for constructing MPSI. In IGOPRF, the receiver obtains the sender-consistent pseudorandom value for an element only when that element is present in every party's set. Otherwise, the receiver obtains an independent random value. This gives MPSI a simple construction approach built around a single primitive, much like the OPRF-based approach for two-party PSI: once IGOPRF is available, it provides a natural basis for building an MPSI protocol. We call this the IGOPRF paradigm.

\begin{table*}[htbp]
\centering
\caption{A comprehensive comparison of MPSI protocols, where $n$ is the input size, $N$ is the number of parties, and $t$ is the maximum number of corrupted parties. The parameters $\lambda$ and $\kappa$ denote the statistical and computational security parameters, respectively. $^{*}$ means that security is proved only for admissible corruption sets, where the designated sender $\mathcal{P}_1$ and the designated receiver $\mathcal{P}_2$ are not corrupted simultaneously. The parameter $\xi$ is a hyperparameter as in~\cite{glov24}. The marker $^{\dagger}$ means that the protocol cannot execute correctly when $t=N-1$ without an additional generic MPC protocol or a trusted third party.}
\resizebox{\textwidth}{!}{
\begin{tabular}{lccccccc}
\toprule
\multirow{2}{*}{\textbf{Protocol}} & \multirow{2}{*}{\textbf{Communication}} & \multirow{2}{*}{\textbf{Computation}} & \multirow{2}{*}{\textbf{Rounds}} & \textbf{Corruption} & \textbf{Security} & \textbf{Single} & \textbf{Functionality} \\
 & & & & \textbf{Bound} & \textbf{Model} & \textbf{Primitive} & \textbf{Extension} \\
\midrule
\multirow{2}{*}{KMP \cite{kmp17}} &$\mathcal{O}(tnN(\lambda + \kappa + \log n))$ &$\mathcal{O}(ntN\kappa)$ & $\mathcal{O}(1)$ &$t<N$ &semi-honest & \textcolor{red}{\xmark} & \textcolor{red}{\xmark} \\
\cmidrule{2-8}
 &$\mathcal{O}(nN(\lambda + \kappa + \log n))$ &$\mathcal{O}(nN\kappa)$ & $\mathcal{O}(1)$ &$t<N$ &augmented semi-honest & \textcolor{red}{\xmark} & \textcolor{red}{\xmark} \\
\midrule
CDG \cite{cdg21} &$\mathcal{O}(nN(\lambda + \kappa + \log n))$ &$\mathcal{O}(nN\kappa)$  & $\mathcal{O}(1)$ &$t < \lfloor (N+1)/2 \rfloor$ &semi-honest & \textcolor{red}{\xmark} & \textcolor{green}{\cmark} \\
\midrule
\multirow{2}{*}{NTY \cite{nty21}} &$\mathcal{O}((N - t)t\kappa + Nn\lambda + tn(\lambda + \kappa + \log n))$ &$\mathcal{O}(Nn\lambda)$& $\mathcal{O}(1)$ &$t<N$ &semi-honest & \textcolor{red}{\xmark} & \textcolor{red}{\xmark} \\
\cmidrule{2-8}
 &$\mathcal{O}((N - t)t\kappa + (N + t)n\kappa + tn\log n)$&$\mathcal{O}(Nn\kappa)$ & $\mathcal{O}(1)$ &$t<N$ &malicious & \textcolor{red}{\xmark} & \textcolor{red}{\xmark} \\
\midrule
\multirow{2}{*}{GLOV \cite{glov24}} &$\mathcal{O}(Nn\log n)$ &$\mathcal{O}(\xi n^{1+1/\xi}+Nn)$ &$\mathcal{O}(\xi)$ &$t<N$ &semi-honest & \textcolor{red}{\xmark} & \textcolor{green}{\cmark} \\
\cmidrule{2-8}
 &$\mathcal{O}(Nn\log n\log\log n+\xi Nn^{1+1/\xi})$ &$\mathcal{O}(Nn\cdot \log n)$ &$\mathcal{O}(\xi)$ &$t<N$ &semi-honest & \textcolor{red}{\xmark} & \textcolor{green}{\cmark} \\
\midrule
WYC \cite{wyc24} &$\mathcal{O}(tNn(\lambda + \kappa + \log n))$&$\mathcal{O}(tNn\kappa)$& $\mathcal{O}(tN)$ &$t<N$ &semi-honest & \textcolor{red}{\xmark} & \textcolor{red}{\xmark} \\
\midrule
\multirow{2}{*}{GLW \cite{glw24}} &$\mathcal{O}(N\kappa + Nn\lambda + \lambda \log n)$ &$\mathcal{O}(Nn\lambda)$ &$\mathcal{O}(1)$ &$(t<N)^{*}$ &semi-honest & \textcolor{red}{\xmark} & \textcolor{green}{\cmark} \\
\cmidrule{2-8}
 &$\mathcal{O}((N + Nn)\kappa + \lambda \log n)$ &$\mathcal{O}(Nn\kappa)$ &$\mathcal{O}(1)$ &$(t<N-1)^{*}$ &malicious clients & \textcolor{red}{\xmark} & \textcolor{green}{\cmark} \\
\midrule
GGA \cite{gga25} &$\mathcal{O}((N + t)n\lambda)$ &$\mathcal{O}((N + t)n\lambda)$  &$\mathcal{O}(1)$ &$t<N$$^{\dagger}$ &semi-honest & \textcolor{red}{\xmark} & \textcolor{red}{\xmark} \\
\midrule
DCZ \cite{dcz25} &$\mathcal{O}(Nn(\lambda + \log n))$ &$\mathcal{O}(nN\lambda)$ &$\mathcal{O}(1)$ &$t<N$ &semi-honest & \textcolor{red}{\xmark} & \textcolor{green}{\cmark} \\
\midrule
\multirow{3}{*}{Ours} &$\mathcal{O}(Nn(\lambda + \log n))$ &$\mathcal{O}(nN\lambda)$ &$\mathcal{O}(1)$ &$(t<N)^{*}$ &semi-honest & \textcolor{green}{\cmark} & \textcolor{green}{\cmark} \\
\cmidrule{2-8}
 &$\mathcal{O}((N+t)n(\lambda + \log n))$ &$\mathcal{O}((N + t)n\lambda)$ &$\mathcal{O}(1)$ &$(t<N)^{*}$ &semi-honest & \textcolor{green}{\cmark} & \textcolor{green}{\cmark} \\
\cmidrule{2-8}
 &$\mathcal{O}((N + t)n\kappa)$ &$\mathcal{O}((N + t)n\kappa)$ &$\mathcal{O}(1)$ &$(t<N)^{*}$ &malicious & \textcolor{green}{\cmark} & \textcolor{green}{\cmark} \\
\bottomrule
\end{tabular}
}
\label{workcomparison}
\end{table*}

Building on this paradigm, we construct MPSI protocols for both the semi-honest and malicious settings under the designated sender-receiver non-collusion assumption also adopted in \cite{glw24}. More specifically, our security model allows any corruption set that does not contain both the designated sender $\mathcal{P}_1$ and the receiver $\mathcal{P}_2$. Under this model, our maliciously secure protocol covers all admissible corruption patterns, whereas \cite{glw24} considers malicious corruption of clients only. Beyond basic MPSI, we also show that IGOPRF can be naturally extended to support other MPSI-related functionalities. Finally, our experiments show that our constructions offer a good balance of security, efficiency, and extensibility compared with existing MPSI protocols.

\subsection{Our Contributions}

We summarize the main contributions of this work as follows.
\begin{itemize}
  \item We introduce a new cryptographic primitive, called IGOPRF, and use it to establish a single-primitive paradigm for constructing MPSI. With IGOPRF in place, constructing MPSI becomes as straightforward as building two-party PSI from OPRF.

  \item We provide concrete instantiations of IGOPRF and use them to construct MPSI protocols in both the semi-honest and malicious settings. Under the designated sender-receiver non-collusion assumption, our maliciously secure protocol covers all admissible corruption patterns. We further extend IGOPRF to an intersection-gated oblivious programmable pseudorandom function (IGOPPRF), which provides a unified way to construct several MPSI variants, including MPSI cardinality, threshold MPSI, and MPSI-sum.

  \item We evaluate our protocols against representative existing MPSI protocols \cite{nty21,wyc24,glw24,gga25,dcz25}, comparing their performance under the respective security models. Table~\ref{exp0} summarizes the online performance in LAN and WAN settings. Compared with NTY \cite{nty21}, our maliciously secure protocol achieves speedups of up to $23.0\times$ in LAN and $6.4\times$ in WAN, while reducing communication by up to a factor of $3.9$. In the semi-honest setting, our protocol is up to $22.0\times$ faster than WYC \cite{wyc24} and $15.8\times$ faster than DCZ \cite{dcz25} in LAN. The largest gains are observed in the malicious setting and in scenarios with large inputs or high corruption tolerance.
\end{itemize}

\begin{table}[htb]
\centering
\caption{Baseline-to-ours runtime and communication ratios under the corresponding evaluated security models. A ratio larger than $1$ indicates that our protocol is faster or communicates less.}
\label{exp0}
\resizebox{0.47\textwidth}{!}{
\begin{tabular}{lccccc}
\toprule
\multirow{2}{*}{\textbf{Baselines}} & \multicolumn{3}{c}{\textbf{Running Time}} & \multirow{2}{*}{\textbf{Communication}} & \multirow{2}{*}{\textbf{Security Model}} \\
\cmidrule(lr){2-4}
 & \textbf{LAN} & \textbf{WAN (400 Mbps)} & \textbf{WAN (200 Mbps)} & & \\
\midrule
NTY~\cite{nty21} & $0.1$--$23.0\times$  & $<0.1$--$3.2\times$ & $<0.1$--$6.4\times$ & $0.2$--$3.9\times$  & Malicious \\
WYC~\cite{wyc24} & $0.1$--$22.0\times$  & $0.1$--$3.8\times$ & $0.1$--$4.1\times$ & $0.3$--$24.7\times$ & Semi-honest \\
GLW~\cite{glw24} & $0.4$--$1.0\times$   & $0.1$--$0.4\times$ & $0.2$--$0.5\times$ & $0.4$--$1.0\times$  & Semi-honest \\
GGA~\cite{gga25} & $<0.1$--$4.5\times$  & $<0.1$--$1.0\times$ & $<0.1$--$1.0\times$ & $0.9$--$1.9\times$  & Semi-honest \\
DCZ~\cite{dcz25} & $0.1$--$15.8\times$ & $<0.1$--$3.6\times$  & $0.1$--$3.1\times$  & $0.8$--$2.2\times$  & Semi-honest \\
\bottomrule
\end{tabular}
}
\end{table}

\section{Technique Overview}
\label{overview}

Our construction builds on the label test used in efficient OPRF-based two-party PSI, with an additional mechanism to extend it to an all-party membership gate. Parties $\mathcal{P}_1$ and $\mathcal{P}_2$ serve as the designated sender and receiver, respectively. Each of the remaining parties contributes a membership condition without learning the receiver's queries or the resulting intersection. Our security model allows any coalition of up to $t$ corrupted parties, provided that it does not contain both designated parties.

\noindent \textbf{From OPRF labels to an all-party gate.}
An OPRF invocation provides access to a pseudorandom function (PRF) $\mathsf{F}^{\mathsf{o}}$. The receiver chooses a single ordered batch of inputs and obtains the corresponding function values. An honest receiver may include at most $q$ inputs in this batch, while a corrupted receiver may include at most $q'$ inputs. The sender can evaluate the same function on its own inputs. The OPRF construction in~\cite{rs21}, which is based on vector oblivious linear evaluation, provides this functionality in the random-oracle model. In our MPSI construction, we set $q=n$ and $q'=n'\geq n$.

In two-party PSI, a sender element and a receiver element match when they produce the same OPRF input and hence receive the same label. Extending this idea to MPSI requires one additional condition: for a receiver element $x$, the sender-side and receiver-side OPRF inputs should match only when $x$ belongs to all $N$ sets. This is exactly what IGOPRF achieves. The receiver obtains $\mathsf{F}^{\mathsf{igo}}(x)$ if $x$ belongs to the intersection of all parties' sets, and an independent label otherwise. Meanwhile, the sender obtains $\mathsf{F}^{\mathsf{igo}}(x_1)$ for its own elements. The sender then pads this label sequence to length $n$ with independent uniform labels and sends it in a random order as $\mathcal{L}_1$. This way, MPSI is reduced to matching labels.

\noindent \textbf{Constructing the gate.}
The key idea is to generate two hidden OPRF inputs that agree only when an element belongs to every party's set. Let $\mathbb{F}$ be a binary extension field, and let $m$ denote the length of the OKVS vectors. Each party first expands a seed using a pseudorandom generator (PRG) to obtain a public vector $S_i\in\mathbb{F}^m$. The parties then combine these vectors to obtain $S=\bigoplus_i S_i$. They also hold additive shares $\Delta_i$ of a nonzero scalar $\Delta$ and sample pads $V_i\in\mathbb{F}^m$. Each party forms $W_i=\Delta_iS\oplus V_i$, and the parties reconstruct $W=\Delta S\oplus V$, where $V=\bigoplus_iV_i$, only to the sender. In this way, $W$ serves as the sender's reference value without revealing the individual shares used to construct it.

However, the receiver should obtain the same value only for elements that belong to every party's set. To achieve this, each party other than the receiver evaluates $S_i$ on every element $x\in X_i$ and encodes the resulting pairs $(x,\mathsf{Decode}(S_i,x))$ into a linear oblivious key-value store (OKVS) vector $T_i$. The receiver simply sets $T_2=S_2$. Now consider an element $x$ in the intersection of all parties' sets. By the correctness of OKVS, we have $\mathsf{Decode}(T_i,x)=\mathsf{Decode}(S_i,x)$ for every party $i$. Therefore, the aggregate vector $T=\bigoplus_iT_i$ decodes to the same value as $S$ at $x$. However, if a receiver query $x$ does not belong to at least one party's set, the OKVS vector $T_i$ of that party decodes to a pseudorandom value at $x$. Consequently, the aggregate vector $T$ also decodes to a pseudorandom value at $x$.

\begin{figure}[htb]
  \centering
  \includegraphics[width=\linewidth]{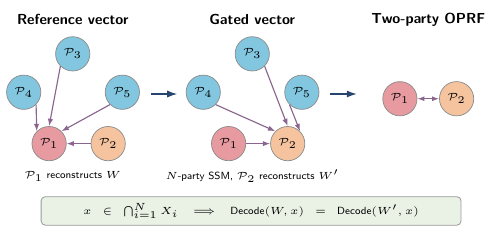}
  \caption{Data flow of $\Pi_{\mathsf{IGOPRF}}^{N-1}$ for $N=5$ and $t=4$.}
  \label{jiagou1}
\end{figure}

The vectors $T_i$ cannot be sent directly to either designated party, as they contain information about each party's local membership. We therefore combine them using shared scalar-vector multiplication (SSM). To do so, we use a vector Beaver triple (VBT)~\cite{beaver91}, which consists of additive shares of a scalar $\Delta$, a random vector $A$, and their product $B=\Delta A$. For each SSM invocation, every participant provides its vector $T_i$, and the protocol returns shares $U_i$ satisfying
$\bigoplus_i U_i=\Delta(\bigoplus_i T_i)$.
The parties then add their respective pads $V_i$ to these shares, allowing the receiver to reconstruct $W'=\Delta T\oplus V$. For any element $x$ in the intersection of all parties' sets, the linearity of OKVS decoding ensures that $\mathsf{Decode}(W',x)=\mathsf{Decode}(W,x)$. The sender and receiver then use their respective decoded values as inputs to the final OPRF. For such an element, the two inputs are equal and therefore produce the same label. For any other receiver query, the receiver's decoded value is pseudorandom and matches the sender's corresponding input only with negligible probability. Figure~\ref{jiagou1} illustrates this data flow.

\noindent \textbf{Reducing the SSM cost to $t+1$ parties.}
The maximal construction runs SSM among all $N$ parties, even when the corruption bound $t$ is much smaller than $N$. However, security against at most $t$ corruptions requires only one honest participant in SSM. Therefore, it is sufficient to run SSM among a core group of $t+1$ parties. We define $S$ and $\Delta$ using only these core parties, while keeping each non-core party's $S_i$ directly in the sender's reference vector. The receiver's gated vector is formed accordingly, with each non-core party contributing its $T_i$ directly.

\begin{figure}[htb]
  \centering
  \includegraphics[width=\linewidth]{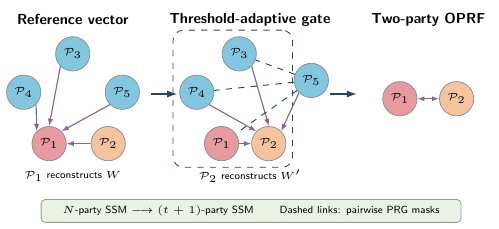}
  \caption{Threshold-adaptive data flow for $N=5$ and $t=3$. The dashed box marks the SSM participants.}
  \label{jiagou2}
\end{figure}

However, directly revealing a non-core party's $T_i$ would expose its local membership information. To avoid this, each non-core party shares a PRG seed with every core party and masks $T_i$ with the XOR of the pads derived from these seeds. Each core party adds the corresponding pads to its SSM output share, so that all pads cancel when the receiver reconstructs $W'$. Since any admissible coalition contains at most $t$ parties, at least one of the $t+1$ core parties remains honest. Thus, every honest non-core party shares at least one pad with an honest core party, keeping its contribution hidden from the coalition. This masking does not change the reconstructed value, so the decoding invariant remains unchanged. As a result, the SSM-dependent work involves only $t+1$ parties instead of all $N$ parties. Figure~\ref{jiagou2} illustrates this optimization.

\noindent \textbf{Binding malicious inputs.}
In the malicious setting, the parties must commit to all randomness, the SSM state, and every set-dependent input before either designated output is released. Each vector $S_i$ is generated from an independently committed-and-opened seed. We use authenticated Beaver correlations to bind $\Delta$, $A$, $B$, and the fresh inputs used in SSM. Let $\langle a\rangle$ denote an authenticated additive sharing of $a$. The parties compute $\langle E\rangle=\langle T\rangle-\langle A\rangle$ and open $E$ with batched verification~\cite{glm23}. They then compute $U_i=B_i+\Delta_iE$. Finally, checked private reconstruction delivers $W$ only to the sender and $W'$ only to the receiver. Any invalid opening, dimension mismatch, authentication failure, or invalid reconstruction causes the protocol to abort.

Malicious security also requires each party's gate to be built from a set within the prescribed size bound. To enforce this, we use a random oracle $\mathsf{H}^{\mathsf{f}}:\{0,1\}^*\to\mathbb{F}$ to map elements into the field. For each non-receiver $\mathcal{P}_i$, the receiver acts as the sender in a separate OPRF instance, with input limits $n$ and $n'$ for honest and corrupted receivers, respectively. Let $\mathsf{F}_i$ denote the PRF used in this instance. For every element $x$ submitted by $\mathcal{P}_i$, the party computes $r_i(x)=\mathsf{F}_i(\mathsf{H}^{\mathsf{f}}(x))$ and encodes $\mathsf{Decode}(S_i,\mathsf{H}^{\mathsf{f}}(x))+r_i(x)$ under the key $\mathsf{H}^{\mathsf{f}}(x)$. The receiver knows the corresponding PRF values for its own elements and uses them to build two correction vectors, $C_{\mathsf{c}}$ and $C_{\mathsf{o}}$. When a non-receiver has submitted an element from the receiver's set, subtracting the corresponding correction cancels the OPRF mask for that element. If the element was not submitted, the corresponding randomized OKVS encoding still decodes to a pseudorandom value.

Before either $W$ or $W'$ is delivered, every pairwise OPRF batch, correction vector, SSM input, and reconstruction input is fixed. With these values fixed, the simulator can extract the effective inputs of corrupted parties. In particular, it extracts the joint effective set of corrupted non-receivers from their accepted OPRF batches and fixed gate messages, and the effective queries of a corrupted receiver from its final OPRF batch. Therefore, the protocol realizes the explicit-set IGOPRF functionality against every admissible malicious coalition.

\begin{figure}[htb]
  \centering
  \includegraphics[width=0.95\linewidth]{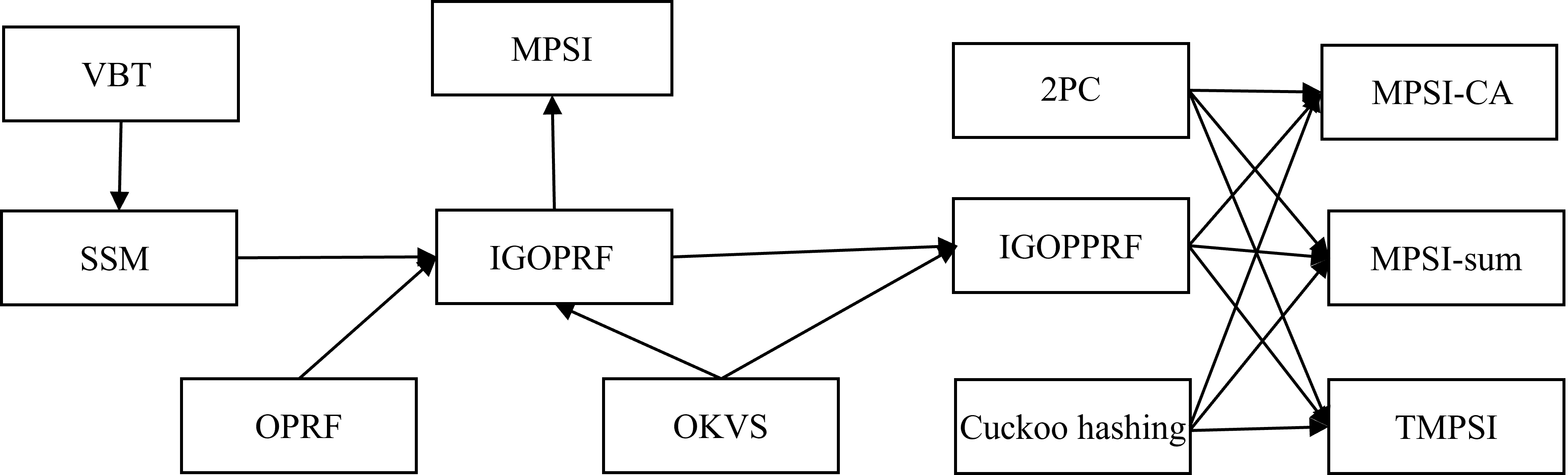}
  \caption{Modular roadmap for MPSI and its variants.}
\label{roadmap}
\end{figure}

\noindent \textbf{Programming the gated output.}
IGOPPRF extends IGOPRF by allowing the sender to associate a payload $z$ with each of its elements. After running IGOPRF, the sender encodes $(x,z\oplus\mathsf{F}^{\mathsf{igo}}(x))$ into an OKVS. For an element in the intersection, the receiver obtains the same IGOPRF label as the sender and can therefore recover the corresponding payload $z$. For any other receiver element, the resulting value remains pseudorandom. We then use Cuckoo hashing~\cite{cuckoo} to align these values for subsequent two-party computation (2PC), which allows us to construct MPSI-cardinality, threshold MPSI, and MPSI-sum. Figure~\ref{roadmap} provides an overview of this construction path.

\section{Related Work}
\label{related-work}

\noindent \textbf{MPSI.}
Early MPSI protocols relied on private polynomial evaluation or additively homomorphic encryption, resulting in quadratic complexity in $n$~\cite{fnp04,ks05}. A more general approach was introduced by Kolesnikov et al.~\cite{kmp17}, who built MPSI from an oblivious programmable pseudorandom function (OPPRF). They proposed two semi-honest protocols with different trade-offs. Their augmented protocol has a cost independent of $t$, but does not tolerate arbitrary collusions. Their standard protocol, in contrast, supports arbitrary collusions. Chandran et al.~\cite{cdg21} later achieved linear complexity under an honest-majority assumption and extended MPSI to support Circuit MPSI and Quorum MPSI. For malicious security, Nevo et al.~\cite{nty21} combined OKVS~\cite{prty20} with recursive scale reduction, obtaining a maliciously secure protocol without tolerance for arbitrary collusions. Wu et al.~\cite{wyc24} proposed two semi-honest protocols based on OPRF and OKVS. These protocols perform well for moderate values of $N$, but their cost scales less favorably with $N$ than that of~\cite{nty21}. Under a designated sender-receiver non-collusion assumption, Gao et al.~\cite{glw24} further achieved efficient semi-honest MPSI and supported MPSI-cardinality. They also considered malicious clients, but their malicious-security guarantee does not cover corrupted designated parties. Other approaches explore different ways to generate the correlations needed for MPSI. Guan et al.~\cite{gga25} introduced a construction based on reducible zero-sharing. When $t=N-1$, their construction requires generic MPC to generate the necessary correlations, and their evaluation uses a trusted third party for this preprocessing. Dong et al.~\cite{dcz25} later introduced predicative zero-sharing for semi-honest multi-party set operations. Their construction, however, does not benefit from a smaller corruption threshold $t$.

\noindent \textbf{MPSI-Cardinality.}
Debnath et al.~\cite{dsk21} introduced the first dedicated MPSI-cardinality construction, using Bloom filters and additively homomorphic encryption, but did not report concrete performance results. Chen et al.~\cite{cdg22} later presented the first practical protocol that tolerates arbitrary collusions, relying on oblivious transfer and symmetric-key operations. Gao et al.~\cite{gty24} explored both server-aided and server-free constructions under a designated-party non-collusion assumption. Meng et al.~\cite{mwl24} subsequently proposed a variant that removes the OPPRF component. The framework of Dong et al.~\cite{dcz25} also supports MPSI-cardinality.

\noindent \textbf{Threshold MPSI.}
Ghosh et al.~\cite{gn19,gs19} presented the first threshold MPSI constructions, achieving linear complexity in $n$ and quadratic complexity in $N$, but without a concrete performance evaluation. Badrinarayanan et al.~\cite{bmr21} later explored a construction based on fully homomorphic encryption. Branco et al.~\cite{bdp21} proposed an approach based on additively homomorphic encryption, with complexity independent of $n$ but quadratic in the intersection threshold $\tau$. More recently, Liu et al.~\cite{lzq23} combined OPRF with secret sharing to construct a probabilistic threshold MPSI protocol. Their protocol, however, outputs the intersection with a prescribed probability for certain ranges of possible intersection sizes.

\noindent \textbf{MPSI-Sum.}
Chen et al.~\cite{cdg22} presented the first practical MPSI-Sum construction based on symmetric-key operations, which reveals the cardinality of the intersection. Dong et al.~\cite{dcz25} later developed a more efficient construction within their general framework, but it incurs the same leakage.

\section{Preliminaries}

\subsection{Notations}

For positive integers $a$ and $b$, we write $[a]=\{1,2,\ldots,a\}$ and $[a,b]=\{a,a+1,\ldots,b\}$ when $a\leq b$, and let $[a,b]=\emptyset$ otherwise. An empty sum or XOR sum is defined to be zero. We use $\kappa$ and $\lambda$ to denote the computational and statistical security parameters, respectively. We consider $N$ parties, where $\mathcal{P}_1$ is the sender, $\mathcal{P}_2$ is the receiver, and the remaining parties $\mathcal{P}_3,\ldots,\mathcal{P}_N$ are clients. Each honest party $\mathcal{P}_i$ holds a set $X_i$. For simplicity, we assume that all honest parties hold sets of the same size $n$. We use $n'\geq n$ as an upper bound on the effective input size of a corrupted party. Let $t$ denote the corruption bound. The admissible corruption sets are defined as $\mathsf{Adm}_t=\{\mathcal{C}\subseteq[N]\mid |\mathcal{C}|\leq t \text{ and } \{1,2\}\not\subseteq\mathcal{C}\}$. Finally, let $\mathbb{F}$ be a binary extension field of size $|\mathbb{F}|=2^{\mathsf{out}}$, where $\mathsf{out}$ is the PRF output length. We use $+$ and $-$ for addition and subtraction in $\mathbb{F}$, respectively, and write $\oplus$ for field addition when emphasizing its XOR representation. The same notation applies componentwise to vectors.

\subsection{Oblivious Key-Value Stores}
\label{okvs}

An OKVS over a key domain $\mathcal{K}$ and a value domain $\mathcal{V}$ encodes $n$ key-value pairs~\cite{rs21,rr22,prty20,prty21,bpsy23,hkls24,tdsc25}. The encoding algorithm $\mathsf{Encode}(\{(k_i,v_i)\}_{i\in[n]})$ outputs a vector $P\in\mathcal{V}^m$ or $\perp$, while the decoding algorithm $\mathsf{Decode}(P,k)$ returns a value in $\mathcal{V}$. Given a key set $K$ and a value vector $v=(v_k)_{k\in K}$ indexed by $K$, we write $\mathsf{Encode}(K,v)$ as a shorthand for the same encoding operation. Correctness requires that $\mathsf{Decode}(P,k_i)=v_i$ for every encoded key $k_i$. The probability of encoding failure is at most $2^{-\lambda}$. Computational obliviousness ensures that the encoded keys remain hidden when the values are uniformly random. We use a linear OKVS, for which decoding preserves addition and scalar multiplication: $\mathsf{Decode}(P+P',k)=\mathsf{Decode}(P,k)+\mathsf{Decode}(P',k)$ and $\mathsf{Decode}(\Delta P,k)=\Delta\mathsf{Decode}(P,k)$. 

\subsection{Vector Beaver Triples}
\label{sec:vbt}

To multiply a shared scalar by a shared vector, the parties use a VBT as defined in Figure~\ref{figvbt}~\cite{beaver91}.

\begin{figure}[htb]
\centering
\fbox{\begin{minipage}{\dimexpr\linewidth-2\fboxsep-2\fboxrule\relax}
\textbf{Parameters:} Participating parties $(\mathcal P_i)_{i\in J}$, field $\mathbb F$, and vector length $m$.
\par\smallskip
\textbf{Functionality:} Sample $\Delta\leftarrow\mathbb F^*$ and $A\leftarrow\mathbb F^m$ independently and set $B=\Delta A$. Choose uniform additive shares satisfying $\sum_{i\in J}(\Delta_i,A_i,B_i)=(\Delta,A,B)$. Privately send $(\Delta_i,A_i,B_i)$ to each $\mathcal P_i$, using authenticated shares in the malicious protocol.
\end{minipage}}
\caption{Ideal functionality $\mathcal F_{\mathsf{VBT}}$ for VBT.}
\label{figvbt}
\end{figure}

\subsection{Oblivious Pseudorandom Functions}
\label{sec:oprf}

Figure~\ref{figoprf} defines the OPRF functionality, which allows the receiver to obtain PRF evaluations on its private inputs~\cite{rs21}.

\begin{figure}[htb]
\centering
\fbox{\begin{minipage}{\dimexpr\linewidth-2\fboxsep-2\fboxrule\relax}
\textbf{Parameters:} A sender and a receiver, field $\mathbb F$, output length $\mathsf{out}$, and receiver bounds $q\le q'$.
\par\smallskip
\textbf{Functionality:} Sample an independent random function $\mathsf F^{\mathsf{o}}:\mathbb F\to\{0,1\}^{\mathsf{out}}$. Accept one receiver batch $(u_1,\ldots,u_s)$, with $s\le q$ if honest and $s\le q'$ if corrupted. Privately return $(\mathsf F^{\mathsf{o}}(u_j))_{j=1}^s$ in order. On each subsequent sender input $u\in\mathbb F$, return $\mathsf F^{\mathsf{o}}(u)$ to the sender.
\end{minipage}}
\caption{Ideal functionality $\mathcal F_{\mathsf{OPRF}}$ for a batch OPRF.}
\label{figoprf}
\end{figure}

\section{Intersection-Gated Oblivious Pseudorandom Function (IGOPRF)}

In this section, we first define IGOPRF and present three concrete constructions: the maximal-collusion, threshold-adaptive, and maliciously secure constructions. We then extend IGOPRF to its programmable variant, IGOPPRF.

\subsection{The Ideal Functionality of IGOPRF}
\label{igoprf}

The IGOPRF functionality in Figure~\ref{figigoprf} specifies how the parties' inputs are fixed and how the corresponding labels are generated. For MPSI, we set the input bounds to $q=n$ and $q'=n'$.

\begin{figure}[htb]
\centering
\fbox{
\parbox{0.95\linewidth}{

\textbf{Parameters:} $N\geq 3$ parties with sender $\mathcal{P}_1$ and receiver $\mathcal{P}_2$, corruption bound $t\in[N-1]$, input bounds $q\leq q'$, and label length $\mathsf{out}$.

\textbf{Functionality:}
For a corruption set $\mathcal{C}\in\mathsf{Adm}_t$, set $X_i^\star=X_i$ for each honest party, where $|X_i|\leq q$. Sample a PRF $\mathsf{F}^{\mathsf{igo}}:\{0,1\}^*\to\{0,1\}^{\mathsf{out}}$.
\par
Receive and fix an ordered query batch from $\mathcal{P}_2$ containing at most $q$ queries if honest and $q'$ queries if corrupted. Each corrupted party then submits its set $X_i^\star$. Reject if any set is missing or exceeds size $q'$. Otherwise, fix all submitted sets.
\par
The adversary may abort before any honest party receives its output. Otherwise, for each query $x$ in the batch, return $\mathsf{F}^{\mathsf{igo}}(x)$ to $\mathcal{P}_2$ if $x\in\bigcap_i X_i^\star$, and an independent uniform $\mathsf{out}$-bit label otherwise. Preserve the query order and reuse the same label for repeated queries.
\par
On input $x_1$ from $\mathcal{P}_1$, return $\mathsf{F}^{\mathsf{igo}}(x_1)$ to $\mathcal{P}_1$.
}
}
\caption{IGOPRF functionality $\mathcal{F}_{\mathsf{IGOPRF}}^t$.}
\label{figigoprf}
\end{figure}

\subsection{IGOPRF Against Maximal Admissible Collusion}
\label{sec:igoprf-max}

We first present $\Pi_{\mathsf{IGOPRF}}^{N-1}$, a semi-honest construction that tolerates every admissible coalition $\mathcal{C}\in\mathsf{Adm}_{N-1}$. The protocol in Figure~\ref{igoprf1} follows the three stages described in Section~\ref{overview}. In Steps~\ref{step:igoprf-max-vbt}--\ref{step:igoprf-max-reference}, the parties construct the sender's reference vector $W$. They then encode their membership conditions and combine the resulting vectors to obtain the receiver's gated vector $W'$ in Steps~\ref{step:igoprf-max-encode}--\ref{step:igoprf-max-reconstruct}. Finally, in Step~\ref{step:igoprf-max-oprf}, the sender and receiver decode their respective vectors and use the resulting values as inputs to the final OPRF.

\begin{figure}[htb]
\centering
\fbox{%
\parbox{0.95\linewidth}{%
\textbf{Inputs:} Each $\mathcal P_i$ holds a set $X_i$. Let $\mathsf{PRG}:\{0,1\}^{\kappa}\to\mathbb F^m$ and $\mathsf{out}=\lambda+\lceil 2\log_2 n\rceil$.
\par\smallskip
\textbf{Protocol:}
\begin{enumerate}
    \item\label{step:igoprf-max-vbt} Each $\mathcal P_i$ receives $(\Delta_i,A_i,B_i)$ from $\mathcal F_{\mathsf{VBT}}$.

    \item $\mathcal P_1$ samples independent seeds $s_i\leftarrow\{0,1\}^{\kappa}$ for all $i\in[N]$ and broadcasts them.

    \item All parties compute $S_i=\mathsf{PRG}(s_i)$ for all $i\in[N]$ and set $S=\bigoplus_{i=1}^{N}S_i$.

    \item Each $\mathcal P_i$ samples $V_i\leftarrow\mathbb F^m$ and computes $W_i=\Delta_iS\oplus V_i$.

    \item\label{step:igoprf-max-reference}\label{step1} Each $\mathcal P_i$ with $i\ne1$ sends $W_i$ to $\mathcal P_1$, which computes $W=\bigoplus_{i=1}^{N}W_i$.

    \item\label{step:igoprf-max-encode} Each $\mathcal P_i$ with $i\ne2$ computes $T_i=\mathsf{Encode}(\{(x,\mathsf{Decode}(S_i,x))\}_{x\in X_i})$. Party $\mathcal P_2$ sets $T_2=S_2$. Abort if encoding fails.

    \item The parties invoke $\mathcal F_{\mathsf{SSM}}$ using the VBT from Step~\ref{step:igoprf-max-vbt}. Each $\mathcal P_i$ inputs $T_i$ and receives $U_i$.

    \item\label{step:igoprf-max-reconstruct} Each $\mathcal P_i$ computes $W_i'=U_i\oplus V_i$. Each $\mathcal P_i$ with $i\ne2$ sends $W_i'$ to $\mathcal P_2$, which computes $W'=\bigoplus_{i=1}^{N}W_i'$.

    \item\label{step:igoprf-max-oprf} The parties $\mathcal P_1$ and $\mathcal P_2$ invoke $\mathcal F_{\mathsf{OPRF}}$ once. For each $x_2\in X_2$, $\mathcal P_2$ obtains $x_2'=\mathsf F^{\mathsf{o}}(\mathsf{Decode}(W',x_2))$. For each $x_1\in X_1$, $\mathcal P_1$ outputs $\mathsf F^{\mathsf{igo}}(x_1)=\mathsf F^{\mathsf{o}}(\mathsf{Decode}(W,x_1))$.
\end{enumerate}
}}
\caption{Semi-honest IGOPRF protocol $\Pi_{\mathsf{IGOPRF}}^{N-1}$.}
\label{igoprf1}
\end{figure}

Each party other than the sender masks $\Delta_i S$ with $V_i$ before sending $W_i$ to the sender, while $\mathcal{F}_{\mathsf{SSM}}$ keeps the individual $T_i$ values hidden during the computation of $W^\prime$. Let $\Delta=\bigoplus_{i=1}^{N}\Delta_i$, $T=\bigoplus_{i=1}^{N}T_i$, and $V=\bigoplus_{i=1}^{N}V_i$. For any element $x_2\in\bigcap_{i=1}^{N}X_i$, the correctness and linearity of OKVS give
\begin{equation*}
\begin{split}
    \mathsf{Decode}(W^\prime,x_2)
    &= \mathsf{Decode}(\Delta T\oplus V,x_2) \\
    &= \mathsf{Decode}(\Delta S\oplus V,x_2) \\
    &= \mathsf{Decode}(W,x_2).
\end{split}
\end{equation*}
Therefore, the sender and receiver use the same OPRF input for every element in the intersection.

Our IGOPRF constructions use the randomized linear OKVS of~\cite{rr22} in its doubly oblivious mode, where the decoding coefficients are determined by the key and public randomness. Given a key set $K$ and the corresponding values $v=(v_k)_{k\in K}$, a successful encoding is statistically close, up to a negligible distance, to a vector sampled uniformly from all vectors $P$ satisfying $\mathsf{Decode}(P,k)=v_k$ for every $k\in K$. We assume that, except with negligible probability, fixing the decoded values on $K$ still allows the decoding at every queried $x\notin K$ to take any value in $\mathbb{F}$, and that distinct evaluated keys have different decoding coefficient vectors.

\begin{theorem}
In the $(\mathcal{F}_{\mathsf{VBT}},\mathcal{F}_{\mathsf{SSM}},\mathcal{F}_{\mathsf{OPRF}})$-hybrid model, the protocol $\Pi_{\mathsf{IGOPRF}}^{N-1}$ in Figure~\ref{igoprf1} realizes the functionality $\mathcal{F}_{\mathsf{IGOPRF}}^{N-1}$ in Figure~\ref{figigoprf} with $q=q'=n$ against a semi-honest adversary.
\end{theorem}

\begin{proof}
Consider a receiver element $x_2\in X_2$ that does not belong to $X_j$ for some $j\ne2$. Fix $S_j$ and consider any two field values $a,b\in\mathbb F$. By the assumed OKVS property, there exist valid encoding vectors that decode to $a$ and $b$ at $x_2$, respectively. Subtracting the first vector from the second gives a vector that decodes to zero on every element in $X_j$ and to $b-a$ at $x_2$. Adding this vector establishes a bijection between the valid encodings that decode to $a$ and those that decode to $b$ at $x_2$. Therefore, every field value occurs equally often at $x_2$ among the valid encodings. Since $T_j$ is sampled nearly uniformly from these encodings, $\mathsf{Decode}(T_j-S_j,x_2)$ is statistically close to uniform over $\mathbb F$. This property is preserved when multiplying by the nonzero scalar $\Delta$ and adding the contributions from the remaining parties.

We next consider two distinct elements $x$ and $y$ in the fixed input sets of the sender and receiver. The difference between their OPRF inputs contains the term $\mathsf{Decode}(V,x)-\mathsf{Decode}(V,y)$, where $V=\sum_iV_i$ is uniformly random and independent of the other terms. Since distinct elements have different decoding coefficients except with negligible probability, there is at least one coordinate at which these coefficients differ. The random value of $V$ at this coordinate makes the difference uniformly distributed over $\mathbb F$. Therefore, unintended OPRF input equalities occur only with negligible probability. Replacing the final OPRF's PRF with a random function, equal inputs receive the same label, while distinct inputs receive independent uniform labels. This gives the label distribution required by $\mathcal{F}_{\mathsf{IGOPRF}}^{N-1}$ for the fixed input sets.

We now consider the view of an adversary corrupting an admissible set of parties $\mathcal C\in\mathsf{Adm}_{N-1}$.

If the sender is corrupted, the receiver remains honest, although some clients may also be corrupted. For each honest party $\mathcal P_i$ with $i\ne1$, the sender receives $W_i=\Delta_iS\oplus V_i$. Since $V_i$ is independently sampled and hidden from the coalition, these messages are jointly uniform and can be simulated without knowing the honest parties' sets. The set-dependent vectors $T_i$ are used only within $\mathcal F_{\mathsf{SSM}}$, and the honest parties' resulting masked shares are sent only to the honest receiver. In particular, the receiver sets $T_2=S_2$, which is independent of $X_2$.

If the receiver is corrupted, the sender remains honest, and any other corrupted parties are clients. The public seeds are independent of the parties' input sets and can therefore be sampled directly by the simulator. Each honest non-receiver sends $W_i'=U_i\oplus V_i$ to the receiver. Since the pads $V_i$ are independently sampled and hidden from the coalition, these messages are jointly uniform. The simulator can therefore generate them without knowing the corresponding $T_i$ values, which remain hidden within $\mathcal F_{\mathsf{SSM}}$.

Finally, if both the sender and receiver are honest, the corrupted clients see only the public seeds, their local state, and their prescribed views of the hybrid functionalities. The simulator can reproduce these views using the corrupted parties' inputs and prescribed outputs.

In all three cases, the simulator uses the corresponding hybrid functionalities to simulate the correlated shares and local states. The final OPRF evaluations are coupled with the ideal labels according to the input equality pattern established above. Consequently, the simulated views are computationally indistinguishable from the real views for every admissible corruption set $\mathcal C\in\mathsf{Adm}_{N-1}$.
\end{proof}

\subsection{IGOPRF Against Arbitrary Admissible Collusion}

The protocol $\Pi_{\mathsf{IGOPRF}}^{N-1}$ requires all $N$ parties to participate in SSM, even when the corruption bound $t$ is much smaller than $N$. To avoid this unnecessary cost, we adapt the construction to $t$ while retaining the same three-stage structure. In Steps~\ref{step:igoprf-adaptive-vbt}--\ref{step:igoprf-adaptive-ssm} of Figure~\ref{igoprf2}, the parties construct the sender's reference vector $W$ and run SSM among only the $t+1$ core parties. In Steps~\ref{step:igoprf-adaptive-outside-pads}--\ref{step:igoprf-adaptive-reconstruct}, the remaining parties contribute their vectors using pairwise pads that cancel when the receiver reconstructs $W'$. Finally, in Step~\ref{step:igoprf-adaptive-oprf}, the sender and receiver obtain their labels through OPRF.

\begin{figure}[htb]
\centering
\fbox{%
\parbox{0.95\linewidth}{%
\textbf{Inputs:} Each $\mathcal P_i$ holds a set $X_i$. Let $t\in[N-1]$ and $\mathsf{out}=\lambda+\lceil 2\log_2 n\rceil$.
\par\smallskip
\textbf{Protocol:}
\begin{enumerate}
    \item\label{step:igoprf-adaptive-vbt} Each core party $\mathcal P_i$ with $i\in[t+1]$ receives $(\Delta_i,A_i,B_i)$ from $\mathcal F_{\mathsf{VBT}}$.

    \item $\mathcal P_1$ samples independent seeds $s_i\leftarrow\{0,1\}^{\kappa}$ for all $i\in[N]$ and broadcasts them.

    \item All parties compute $S_i=\mathsf{PRG}(s_i)$ for all $i\in[N]$ and set $S=\bigoplus_{i=1}^{t+1}S_i$.

    \item Each core party $\mathcal P_i$ samples $V_i\leftarrow\mathbb F^m$ and computes $W_i=\Delta_i S\oplus V_i$. Each outside party $\mathcal P_i$ sets $W_i=S_i$.

    \item Each $\mathcal P_i$ with $i\ne1$ sends $W_i$ to $\mathcal P_1$, which computes $W=\bigoplus_{i=1}^{N}W_i$.

    \item Each $\mathcal P_i$ with $i\ne2$ computes $T_i=\mathsf{Encode}(\{(x,\mathsf{Decode}(S_i,x))\}_{x\in X_i})$. Party $\mathcal P_2$ sets $T_2=S_2$. Abort if encoding fails.

    \item\label{step:igoprf-adaptive-ssm} The core parties invoke $\mathcal F_{\mathsf{SSM}}$ using the VBT from Step~\ref{step:igoprf-adaptive-vbt}. Each core party $\mathcal P_i$ inputs $T_i$ and receives $U_i$.

    \item\label{step:igoprf-adaptive-outside-pads} Each outside party $\mathcal P_i$ samples an independent seed $s_{i,j}\leftarrow\{0,1\}^{\kappa}$ for each core party $\mathcal P_j$ and privately sends it to $\mathcal P_j$. It then computes $W_i'=T_i\oplus\bigoplus_{j=1}^{t+1}\mathsf{PRG}(s_{i,j})$.

    \item\label{step:igoprf-adaptive-reconstruct} Each core party $\mathcal P_i$ computes $W_i'=U_i\oplus V_i\oplus\bigoplus_{j\in[t+2,N]}\mathsf{PRG}(s_{j,i})$. Each $\mathcal P_i$ with $i\ne2$ sends $W_i'$ to $\mathcal P_2$, which computes $W'=\bigoplus_{i=1}^{N}W_i'$.

    \item\label{step:igoprf-adaptive-oprf} Using $W$ and $W'$ respectively, $\mathcal P_1$ and $\mathcal P_2$ obtain their labels as in Step~\ref{step:igoprf-max-oprf} of Figure~\ref{igoprf1}.
\end{enumerate}
}}
\caption{Semi-honest IGOPRF protocol $\Pi_{\mathsf{IGOPRF}}^{t}$ against arbitrary admissible collusions.}
\label{igoprf2}
\end{figure}

In Figure~\ref{igoprf2}, only the first $t+1$ parties $\mathcal{P}_1,\ldots,\mathcal{P}_{t+1}$ participate in $\mathcal{F}_{\mathsf{SSM}}$. Each outside party masks its vector $T_i$ using PRG pads shared with these core parties. Since at most $t$ parties can be corrupted, at least one core party remains honest. Therefore, for every honest outside party, at least one of its pads remains unknown to the adversary, keeping its vector $T_i$ hidden. By limiting SSM to the core parties, we reduce the SSM-dependent cost from $\mathcal{O}(Nn\lambda)$ to $\mathcal{O}(tn\lambda)$.

To see why the construction is correct, consider an element $x_2\in\bigcap_{i=1}^{N}X_i$. Let $\Delta=\bigoplus_{i=1}^{t+1}\Delta_i$ and $V=\bigoplus_{i=1}^{t+1}V_i$. Since all pairwise pads cancel during the reconstruction of $W^\prime$, the correctness and linearity of OKVS give
\begin{equation*}
\begin{aligned}
 \mathsf{Decode}(W^\prime,x_2)
 &= \mathsf{Decode}\bigg(
      \Delta\bigg(\bigoplus_{i=1}^{t+1}T_i\bigg)
      \oplus V \\
 &\qquad\oplus\bigg(\bigoplus_{i\in[t+2,N]}T_i\bigg),x_2\bigg) \\
 &= \mathsf{Decode}\bigg(
      \Delta\bigg(\bigoplus_{i=1}^{t+1}S_i\bigg)
      \oplus V \\
 &\qquad\oplus\bigg(\bigoplus_{i\in[t+2,N]}S_i\bigg),x_2\bigg) \\
 &= \mathsf{Decode}(W,x_2).
\end{aligned}
\end{equation*}
Therefore, the sender and receiver use the same OPRF input for every element in the intersection.

\begin{theorem}
In the $(\mathcal{F}_{\mathsf{VBT}},\mathcal{F}_{\mathsf{SSM}},\mathcal{F}_{\mathsf{OPRF}})$-hybrid model, $\Pi_\mathsf{IGOPRF}^{t}$ in Figure~\ref{igoprf2} realizes $\mathcal{F}_{\mathsf{IGOPRF}}^{t}$ in Figure~\ref{figigoprf} with $q=q'=n$ against a semi-honest adversary.
\end{theorem}

\begin{proof}
The correctness argument above shows that the sender and receiver obtain the same OPRF input for every element in the all-party intersection. Now consider a receiver query $x_2\in X_2$ that does not belong to some non-receiver's set $X_j$. By the argument in Section~\ref{sec:igoprf-max}, $\mathsf{Decode}(T_j-S_j,x_2)$ is statistically close to uniform over $\mathbb F$. If $\mathcal P_j$ is a core party, its contribution to the difference between the receiver's and sender's OPRF inputs is multiplied by $\Delta$. Otherwise, the coefficient is one. Since $\Delta$ is nonzero, both cases preserve uniformity. After fixing the contributions from all other parties, the difference between the two OPRF inputs remains statistically close to uniform. Therefore, the receiver's input matches the sender's input only with negligible probability for a non-intersection element.

We also need to ensure that distinct elements produce distinct OPRF inputs. For any two distinct elements $x$ and $y$, the difference between their OPRF inputs contains $\mathsf{Decode}(V,x)-\mathsf{Decode}(V,y)$, where $V=\bigoplus_{i=1}^{t+1}V_i$ is uniformly random. By the same argument as in the maximal-collusion construction, the decoding rows of distinct elements are different except with negligible probability. Their difference applied to $V$ is therefore uniform over $\mathbb F$. Thus, unintended OPRF input equalities occur only with negligible probability. The final random-function OPRF consequently assigns the same label to repeated inputs and independent uniform labels to distinct inputs, giving the output distribution required by $\mathcal{F}_{\mathsf{IGOPRF}}^{t}$.

We now consider an arbitrary corruption set $\mathcal C\in\mathsf{Adm}_t$. Since $|\mathcal C|\leq t$, at least one of the $t+1$ core parties remains honest. Moreover, the sender and receiver cannot both be corrupted.

If the sender is corrupted, the receiver remains honest, although some clients may also be corrupted. For every honest core party $\mathcal P_i$, the sender receives $W_i=\Delta_iS\oplus V_i$. Since $V_i$ is independently sampled and hidden from the coalition, this message is uniformly distributed and reveals no information about the honest party's set. An outside party sends only $W_i=S_i$, which is determined by its public seed and is independent of its set. The set-dependent inputs of honest core parties remain hidden within $\mathcal F_{\mathsf{SSM}}$. Each honest outside party also masks its $T_i$ with pairwise pads shared with the core parties. Since at least one core party is honest, at least one of these pads remains unknown to the coalition, keeping $T_i$ hidden.

If the receiver is corrupted, the sender remains honest, and any other corrupted parties are clients. For each honest core party $\mathcal P_i$, the message $W_i'=U_i\oplus V_i\oplus(\bigoplus_{j\in[t+2,N]}\mathsf{PRG}(s_{j,i}))$ is hidden by the fresh random pad $V_i$. Each honest outside party sends $W_i'=T_i\oplus(\bigoplus_{j=1}^{t+1}\mathsf{PRG}(s_{i,j}))$. Since the honest sender $\mathcal P_1$ is a core party, the pad $\mathsf{PRG}(s_{i,1})$ remains unknown to the coalition and hides $T_i$. The final OPRF labels reveal only the outputs prescribed by $\mathcal F_{\mathsf{OPRF}}$.

Finally, suppose both designated parties are honest. The set-dependent vector $T_1$ remains inside $\mathcal F_{\mathsf{SSM}}$, while the receiver uses its private set to form its final OPRF queries and perform local decoding. The corrupted clients see the public seeds, the pairwise seeds sent to them, their own local states, and the messages prescribed by the hybrid functionalities. The simulators for $\mathcal F_{\mathsf{VBT}}$, $\mathcal F_{\mathsf{SSM}}$, and $\mathcal F_{\mathsf{OPRF}}$ reproduce these views using the corrupted parties' inputs and prescribed outputs.

In all three cases, the simulated views are computationally indistinguishable from the real views. Therefore, $\Pi_\mathsf{IGOPRF}^{t}$ realizes $\mathcal{F}_{\mathsf{IGOPRF}}^{t}$ against every admissible semi-honest coalition.
\end{proof}

\subsection{IGOPRF Against Malicious Adversaries}
\label{sec:malicious-igoprf}

We now extend the threshold-adaptive construction to the malicious setting. As in the semi-honest construction, the protocol in Figure~\ref{igoprf3} proceeds in three stages. It first constructs the sender's reference vector, then builds the receiver's gated vector, and finally evaluates the output labels. The first stage (Steps~\ref{step:igoprf-mal-vbt}--\ref{step:igoprf-mal-reference-shares}) establishes the authenticated preprocessing and fixes the public randomness used to construct the reference vector. The second stage (Steps~\ref{step:igoprf-mal-pairwise-oprf}--\ref{step:igoprf-mal-reconstruct}) uses pairwise OPRF calls to bind each non-receiver's contribution to a bounded input set. The receiver also builds the correction vectors needed to remove the OPRF masks. The parties then perform authenticated SSM and fix all relevant inputs before privately reconstructing $W$ and $W'$. In the final stage, Step~\ref{step:igoprf-mal-oprf} invokes the OPRF to generate the output labels under the same input bounds.

\begin{figure*}[t!]
\centering
\fbox{%
\parbox{0.95\linewidth}{%
\textbf{Inputs:} Each $\mathcal P_i$ holds a set $X_i$. Let $t\in[N-1]$, $\mathsf{out}=\kappa+2\log_2 n$, and let $\mathsf H^{\mathsf f}:\{0,1\}^*\to\mathbb F$ be a random oracle. Each OPRF invocation uses receiver input bounds $n$ and $n'$.
\par\smallskip
\textbf{Protocol:} Abort on any invalid opening, failed encoding, malformed vector, or failed authentication or reconstruction check.
\begin{enumerate}
    \item\label{step:igoprf-mal-vbt} Each core party $\mathcal P_i$ with $i\in[t+1]$ receives authenticated shares $(\Delta_i,A_i,B_i)$ from $\mathcal F_{\mathsf{VBT}}$.

    \item Each $\mathcal P_i$ independently samples $s_i\leftarrow\{0,1\}^{\kappa}$. All parties commit to their seeds before opening them.

    \item All parties compute $S_i=\mathsf{PRG}(s_i)$ for $i\in[N]$ and set $S=\bigoplus_{i=1}^{t+1}S_i$.

    \item\label{step:igoprf-mal-reference-shares} The core parties use authenticated random sharing to obtain additive shares $V_i$ of an independent uniform vector $V\in\mathbb F^m$. They fix these shares together with the VBT state and compute $W_i=\Delta_iS\oplus V_i$ using authenticated shares.

    \item\label{step:igoprf-mal-pairwise-oprf} For each $i\ne2$, $\mathcal P_2$ and $\mathcal P_i$ invoke an independent $\mathcal F_{\mathsf{OPRF}}$, with $\mathcal P_2$ as sender and $\mathcal P_i$ as receiver. For each $x\in X_i$, $\mathcal P_i$ obtains $r_i(x)=\mathsf F_i(\mathsf H^{\mathsf f}(x))$, where $\mathsf F_i$ is the PRF used in this invocation.

    \item\label{step:igoprf-mal-gate-encodings} Let $\eta_x=\mathsf H^{\mathsf f}(x)$. Each $\mathcal P_i$ with $i\ne2$ computes $T_i=\mathsf{Encode}(\{(\eta_x,\mathsf{Decode}(S_i,\eta_x)+r_i(x))\}_{x\in X_i})$. For each $x\in X_2$, $\mathcal P_2$ computes
    \[
    c_{\mathsf{c}}(x)=\sum_{i\in[t+1]\setminus\{2\}}\mathsf F_i(\eta_x),
    \qquad
    c_{\mathsf{o}}(x)=\sum_{i\in[t+2,N]}\mathsf F_i(\eta_x).
    \]
    It then computes $C_j=\mathsf{Encode}(\{(\eta_x,c_j(x))\}_{x\in X_2})$ for $j\in\{\mathsf c,\mathsf o\}$ and sets $T_2=S_2-C_{\mathsf c}$.

    \item\label{step:igoprf-mal-ssm} The core parties invoke $\mathcal F_{\mathsf{SSM}}$ using the authenticated triple from Step~\ref{step:igoprf-mal-vbt}. Each core party $\mathcal P_i$ inputs $T_i$ and receives its authenticated share $U_i$.

    \item Each outside party $\mathcal P_i$ samples an independent seed $s'_{i,j}\leftarrow\{0,1\}^{\kappa}$ for each core party $\mathcal P_j$ and privately sends it to $\mathcal P_j$. Both parties derive $p_{i,j}=\mathsf{PRG}(s'_{i,j})$, and each core party submits $p_{i,j}$ as an authenticated input. Each outside party $\mathcal P_i$ computes $W_i'=T_i\oplus\bigoplus_{j=1}^{t+1}p_{i,j}$ and submits it as an authenticated reconstruction input.

    \item\label{step:igoprf-mal-reconstruct} Each core party $\mathcal P_i$ computes $W_i'=U_i\oplus V_i\oplus\bigoplus_{j\in[t+2,N]}p_{j,i}$ using authenticated shares. Before either vector is reconstructed, all pairwise OPRF batches and accepted gate, pad, and reconstruction inputs are fixed. An honest receiver also fixes both correction vectors. Checked private reconstruction delivers $W=\bigoplus_{i=1}^{t+1}W_i\oplus\bigoplus_{i\in[t+2,N]}S_i$ only to $\mathcal P_1$ and $W'=\bigoplus_{i=1}^{N}W_i'$ only to $\mathcal P_2$. The receiver then computes $\widetilde W'=W'-C_{\mathsf o}$.

    \item\label{step:igoprf-mal-oprf} The parties invoke $\mathcal F_{\mathsf{OPRF}}$ once more, with $\mathcal P_1$ as sender and $\mathcal P_2$ as receiver. For each $x_2\in X_2$, the receiver queries $\mathsf{Decode}(\widetilde W',\eta_{x_2})$ and obtains $x_2'=\mathsf F^{\mathsf o}(\mathsf{Decode}(\widetilde W',\eta_{x_2}))$. For each $x_1\in X_1$, the sender outputs $\mathsf F^{\mathsf{igo}}(x_1)=\mathsf F^{\mathsf o}(\mathsf{Decode}(W,\eta_{x_1}))$.
\end{enumerate}
}}
\caption{Maliciously secure IGOPRF protocol $\Pi_\mathsf{IGOPRF}^{t,\mathsf{mal}}$.}
\label{igoprf3}
\end{figure*}

For correctness, let $D_i=T_i-S_i$ and $\eta_x=\mathsf{H}^{\mathsf f}(x)$. For every honest non-receiver $\mathcal P_i$ and every $x\in X_i$, OKVS correctness gives $\mathsf{Decode}(D_i,\eta_x)=\mathsf F_i(\eta_x)$. Let $\Delta=\sum_{i=1}^{t+1}\Delta_i$. For each $x\in X_2$, the linearity of OKVS decoding gives
\begin{multline*}
\mathsf{Decode}(\widetilde{W}^{\prime}-W,\eta_x)\\
=\Delta\left(
\sum_{\substack{i\in[t+1]\\i\ne2}}
\mathsf{Decode}(D_i,\eta_x)
-c_{\mathsf{c}}(x)\right)\\
+\left(
\sum_{i\in[t+2,N]}
\mathsf{Decode}(D_i,\eta_x)
-c_{\mathsf{o}}(x)\right).
\end{multline*}
If $x$ belongs to the intersection of all parties' sets, each gate contribution matches the corresponding PRF value in the correction vectors. Both parenthesized terms therefore vanish, giving $\mathsf{Decode}(\widetilde{W}^{\prime},\eta_x)=\mathsf{Decode}(W,\eta_x)$.

For an element outside a party's accepted OPRF batch, the corresponding correction term still contains $\mathsf F_i(\eta_x)$. For example, if a corrupted party sets $T_i=S_i$, then $\mathsf{Decode}(D_i,\eta_x)=0$, leaving $-\mathsf F_i(\eta_x)$ in the corresponding term. Corrupted parties may coordinate their gate messages to cancel such terms. However, doing so for an element outside their accepted OPRF inputs requires predicting the corresponding PRF value, which succeeds only with negligible probability.

We instantiate malicious SSM using an authenticated Beaver correlation $(\langle\Delta\rangle,\langle A\rangle,\langle B\rangle)$, where $T=\sum_iT_i$ and $B=\Delta A$. We generate this correlation once through authenticated Beaver-triple preprocessing, with $\Delta\in\mathbb F^*$. SSM accepts only fresh authenticated shares of the $T_i$ values under this recorded correlation state. The parties compute $\langle E\rangle=\langle T\rangle-\langle A\rangle$ and publicly open the masked value $E$ using batched verification~\cite{glm23}. The protocol aborts if the verification fails. Each party then computes $U_i=B_i+\Delta_iE$, so that $\sum_iU_i=B+\Delta E=\Delta T$.

We use a single stateful authenticated-sharing mechanism throughout the computation. It supports authenticated random sharing and authenticated external input, and records the accepted SSM inputs, pads, and reconstruction inputs so that they cannot be changed later. Its private checked-reconstruction operation delivers the reconstructed value only to the designated recipient. All other parties learn only whether the reconstruction is accepted or the protocol aborts. We include these operations in the stateful VBT/SSM hybrid used in Theorem~\ref{thm:mal-igoprf}. The public opening of $E$ and the batched message authentication code (MAC) verification follow the technique of~\cite{glm23}. The private delivery of $W$ and $W^\prime$ uses a separate authenticated-output operation. Any failed input check, MAC verification, or reconstruction check causes the protocol to abort.

Prior work provides concrete implementations of the authenticated-sharing and OPRF components~\cite{dklp13,kos16,tdsc251,cm20,rs21,bpsy23,tdsc25}. All pairwise OPRF batches and gate-related authenticated inputs are fixed before either $W$ or $W^\prime$ is reconstructed.

\begin{theorem}
In the $(\mathcal{F}_{\mathsf{VBT}},\mathcal{F}_{\mathsf{SSM}},\mathcal{F}_{\mathsf{OPRF}},\mathsf{H}^{\mathsf{f}})$-hybrid model, $\Pi_\mathsf{IGOPRF}^{t,\mathsf{mal}}$ in Figure~\ref{igoprf3} realizes the explicit-set functionality $\mathcal{F}_{\mathsf{IGOPRF}}^{t}$ with $q=n$ and $q'=n'$ against malicious adversaries for every $\mathcal{C}\in\mathsf{Adm}_t$.
\label{thm:mal-igoprf}
\end{theorem}

\begin{proof}
Fix an adversary $\mathcal{A}$ corrupting $\mathcal{C}\in\mathsf{Adm}_t$. The simulator runs $\mathcal A$ internally, controls $\mathsf{H}^{\mathsf{f}}$, and emulates the hybrid functionalities. Every OPRF invocation has an independent random-function table. In pairwise instance $i$ with a corrupted receiver, the simulator waits for the accepted input batch $\mathcal{B}_i\subseteq\mathbb{F}$ at time $\theta_i$ and extracts
$Y_i=\{x:x\text{ was queried to }\mathsf{H}^{\mathsf{f}}\text{ before }\theta_i\text{ and }\mathsf{H}^{\mathsf{f}}(x)\in\mathcal{B}_i\}$. The simulator fixes $Y_i$ at $\theta_i$ and treats every accepted field input without a recorded preimage as a dummy. With distinct oracle images, $|Y_i|\le n'$. Each subsequent oracle query on a fresh raw input has a uniform image in $\mathbb F$.

Invalid commitment openings, malformed dimensions, failed authenticated operations, and hybrid aborts become ideal aborts. In an accepted execution, the VBT state fixes one $\Delta\in\mathbb{F}^*$ and authenticated sharing fixes an independent uniform $V$. Let $\overline T_i$ be the accepted SSM input of core party $\mathcal P_i$. For $i\in[t+2,N]$, let $\overline W_i^{\prime}$ be its accepted reconstruction input and $\widetilde p_{i,j}$ the pad accepted from core party $\mathcal P_j$. If $\mathcal P_i$ is corrupted, define $\overline T_i=\overline W_i^{\prime}+\sum_{j=1}^{t+1}\widetilde p_{i,j}$. Otherwise, set $\overline T_i=T_i$. The pad discrepancy introduced by corrupted core parties toward honest outside parties is $\Gamma_{\mathcal C}=\sum_{i\in[t+2,N]\setminus\mathcal C}\sum_{j\in\mathcal C\cap[t+1]}(\widetilde p_{i,j}-p_{i,j})$.

An honest core party has $\overline T_i=T_i$, and an honest core recipient accepts the prescribed $p_{i,j}$. Since $\mathbb{F}$ has characteristic two, authenticated linear computation fixes
\[
\begin{aligned}
W&=\Delta S+V+\sum_{i\in[t+2,N]}S_i,\\
W_{\mathsf{rec}}^{\prime}&=\Delta\sum_{i=1}^{t+1}\overline T_i+V
+\sum_{i\in[t+2,N]}\overline T_i+\Gamma_{\mathcal C}.
\end{aligned}
\]
An honest receiver subtracts its already fixed correction vector to obtain $\widetilde W^{\prime}$. A corrupted receiver may use an arbitrary vector. Every pairwise OPRF batch, $\overline T_i$, pad, reconstruction input, and $\Gamma_{\mathcal C}$ is fixed before either aggregate is privately reconstructed. The discrepancy is independent of the randomness each honest party uses to sample its OKVS encoding.

Write $\eta_x=\mathsf{H}^{\mathsf{f}}(x)$, $D_i=\overline T_i-S_i$, and $e_i(x)=\mathsf{Decode}(D_i,\eta_x)-\mathsf{F}_i(\eta_x)$ for $i\ne2$. An honest non-receiver has $e_i(x)=0$ on $X_i$. For an unencoded key $\eta_x$, the decoding argument in Section~\ref{sec:igoprf-max} applies with the pairwise function $\mathsf F_i$ fixed, so $e_i(x)$ is statistically close to uniform. If $\mathcal P_2$ is honest and $i$ is corrupted, the coalition obtains the pairwise values on $Y_i$. For $x\notin Y_i$, the value $\mathsf F_i(\eta_x)$ remains hidden unless $\eta_x$ coincides with an accepted field input.

Write $w_x=\mathsf{Decode}(W,\eta_x)$ and, for an honest receiver, $d_x=\mathsf{Decode}(\widetilde W^\prime,\eta_x)$. The corrupted parties' preprocessing shares and the masked SSM opening can be sampled without honest inputs. Until private reconstruction an honest core share of $V$ hides both aggregates, and admissibility allows the coalition to receive at most one of them. For two fixed candidates with different decoding coefficients, at least one coordinate of the hidden honest share has a nonzero coefficient in the difference of their evaluations. This makes the difference uniform over $\mathbb F$.

When $\mathcal P_2$ is honest and $\mathcal C\ne\emptyset$, put $Y=\bigcap_{i\in\mathcal C}Y_i$. For each $x\in Y$, the simulator computes the contribution of the corrupted parties
\[
\begin{aligned}
e_{\mathcal C}^{\Gamma}(x)
&=\Delta\sum_{i\in\mathcal C\cap[t+1]}e_i(x)
+\sum_{i\in\mathcal C\cap[t+2,N]}e_i(x)\\
&\quad+\mathsf{Decode}(\Gamma_{\mathcal C},\eta_x).
\end{aligned}
\]
If every honest non-receiver holds $x$, the receiver's corrected input satisfies $d_x=w_x+e_{\mathcal C}^{\Gamma}(x)$. The simulator uses this identity differently depending on who receives $W$.

If the sender is corrupted, the simulator submits $Y$ for each corrupted non-receiver and finalizes the gate before releasing $W$. For every $x\in Y$, it computes $d_x^+=w_x+e_{\mathcal C}^{\Gamma}(x)$ from its sampled $W$, the preprocessing state, and the accepted corrupted inputs. Before delivering $W$, it assigns the ideal sender evaluation $\mathsf F^{\mathsf{igo}}(x)$ to the simulated final OPRF entry at $d_x^+$. The uniform pad hides this entry until reconstruction. All other entries are answered lazily, including arbitrary sender-side field queries. If every honest non-receiver holds $x$, then the honest receiver evaluates exactly at $d_x^+$, including when $d_x^+\ne w_x$. If an honest non-receiver omits $x$, its randomized OKVS encoding masks the receiver's actual input. For $x\notin Y$, an unqueried pairwise value masks that input even if every honest non-receiver holds $x$. A receiver input that the sender has not queried gives a label independent of the sender's view. The simulator uses no honest receiver query or output.

If the receiver is corrupted, the simulator samples the privately delivered $W_{\mathsf{rec}}^\prime$ uniformly conditional on the coalition's prior view. The hidden share of $V$ gives this distribution even for fixed honest encodings. For a candidate $x$, set $a_i=\mathsf{Decode}(\overline T_i-S_i,\eta_x)$ for each corrupted party, including the receiver, and $a_i=\mathsf F_i(\eta_x)$ for each honest party. The simulator obtains the latter values because $\mathcal P_2$ is the sender in every pairwise OPRF. It computes
\[
\begin{aligned}
g_x={}&\mathsf{Decode}(W_{\mathsf{rec}}^\prime-\Gamma_{\mathcal C},\eta_x)\\
      &-\Delta\sum_{i=1}^{t+1}a_i-\sum_{i\in[t+2,N]}a_i.
\end{aligned}
\]
Here $a_2$ uses the receiver's actual accepted vector $\overline T_2$. If every honest non-receiver holds $x$, the reconstruction equation gives $g_x=w_x$. If an honest non-receiver omits $x$, the value decoded from its randomized OKVS encoding hides $w_x$ even given $W_{\mathsf{rec}}^\prime$.

At acceptance of the final batch $\mathcal B_{\mathsf{fin}}$, the simulator takes the previously queried raw elements $x$ with $g_x\in\mathcal B_{\mathsf{fin}}$. To see why these elements occupy distinct entries, sample the successive fresh $\mathsf H^{\mathsf f}$ replies in advance. With these field points, the pairwise function tables and the gate inputs fixed, $g_x-g_y$ depends on the uniform vector $W_{\mathsf{rec}}^\prime$ only through the difference between its decodings at $\eta_x$ and $\eta_y$. This difference is uniform for distinct decoding coefficients, so a collision among the presampled candidates is negligible even if the adversary selects a pair after reconstruction. The simulator submits the extracted elements as the receiver's ideal batch and as each corrupted party's effective set, then places the prescribed labels at the corresponding field entries. It samples the remaining entries independently and reuses labels for repeated inputs in the original order. For a raw element first queried after batch acceptance, its oracle image must hit an already evaluated field argument or the term $-\Delta\mathsf F_1(\eta_x)$ in $g_x$ contains a fresh pairwise value at the time the batch was fixed. Here $\mathcal P_1$ is honest and $\Delta\ne0$. Thus a new raw element matches an entry of the fixed batch only with negligible probability. Extraction requires neither the honest sets nor the hidden $W$.

If both designated parties are honest and $\mathcal C\ne\emptyset$, the simulator submits $\{x\in Y:e_{\mathcal C}^{\Gamma}(x)=0\}$ for every corrupted client. Both honest parties now use their prescribed final inputs. For $x\in Y$ held by every honest party these inputs agree exactly when $e_{\mathcal C}^{\Gamma}(x)=0$. An honest omission or a missing pairwise evaluation masks the difference elsewhere. For $\mathcal C=\emptyset$ the honest-execution identity applies directly. The clients receive no final OPRF output in either case.

With polynomially many queries, the stated field and label lengths make unintended matches negligible. Under the assumptions of the theorem, the real and ideal executions give computationally indistinguishable joint distributions of the adversary's view and the parties' outputs.
\end{proof}

\subsection{Extension of IGOPRF to IGOPPRF}
\label{sec:igopprf}

IGOPPRF extends IGOPRF by allowing the sender to associate a payload with each of its elements while retaining the same all-party intersection gate. The ideal functionality in Figure~\ref{figigopprf} specifies this extension. The protocol in Figure~\ref{igopprf} follows a simple approach: it first obtains the gated labels through IGOPRF, then uses these labels to mask the sender's payloads in an OKVS, and finally allows the receiver to recover the payloads for elements in the all-party intersection.

\begin{figure}[htb]
\centering
\fbox{
\parbox{0.95\linewidth}{

\textbf{Parameters:} $N\geq3$ parties, corruption bound $t\in[N-1]$, capacity $q$, and output length $\mathsf{out}$.

\textbf{Functionality:} For $\mathcal C\in\mathsf{Adm}_t$, party $\mathcal P_1$ supplies $\mathcal Z_1=\{(x,z_x):x\in X_1\}$ with $\mathsf{out}$-bit payloads, and each other party $\mathcal P_i$ supplies $X_i$, where $|X_i|\leq q$. Define $\mathsf F^{\mathsf{igop}}(x)=z_x$ for $x\in X_1$ and choose independent pseudorandom values for inputs outside $X_1$.
\par
Fix one ordered query batch of at most $q$ elements from $\mathcal P_2$. For each query $x$, return $\mathsf F^{\mathsf{igop}}(x)$ to $\mathcal P_2$ if $x\in\bigcap_i X_i$. Otherwise, return an independent uniform $\mathsf{out}$-bit value for each distinct query, reusing the same value for repeated queries.
\par
On input $x\in X_1$, return $\mathsf F^{\mathsf{igop}}(x)$ to $\mathcal P_1$.
}
}
\caption{IGOPPRF functionality $\mathcal F_{\mathsf{IGOPPRF}}^t$.}
\label{figigopprf}
\end{figure}

\begin{figure}[htb]
\centering
\fbox{%
\parbox{0.95\linewidth}{%
\textbf{Inputs:} $\mathcal P_1$ holds $\mathcal Z_1=\{(x,z_x):x\in X_1\}$, and every other $\mathcal P_i$ holds $X_i$, where $|X_i|\leq q$ and each payload has $\mathsf{out}$ bits.
\par\smallskip
\textbf{Protocol:}
\begin{enumerate}
    \item Invoke $\mathcal F_{\mathsf{IGOPRF}}^t$ on the sets $X_i$ with $q'=q$. Party $\mathcal P_2$ obtains $x_2'$ for each $x_2\in X_2$.

    \item From the same invocation, $\mathcal P_1$ obtains $\mathsf F^{\mathsf{igo}}(x)$ for each $x\in X_1$.

    \item $\mathcal P_1$ sends $P=\mathsf{Encode}(\{(x,z_x\oplus\mathsf F^{\mathsf{igo}}(x))\}_{x\in X_1})$ to $\mathcal P_2$.

    \item For each $x_2\in X_2$, $\mathcal P_2$ outputs $x_2''=x_2'\oplus\mathsf{Decode}(P,x_2)$. For each $x_1\in X_1$, $\mathcal P_1$ outputs $\mathsf F^{\mathsf{igop}}(x_1)=\mathsf F^{\mathsf{igo}}(x_1)\oplus\mathsf{Decode}(P,x_1)$.
\end{enumerate}
}}
\caption{IGOPPRF protocol $\Pi_{\mathsf{IGOPPRF}}^t$.}
\label{igopprf}
\end{figure}

To see why the protocol works, consider an element $x_2\in\bigcap_{i=1}^{N}X_i$. Since $x_2\in X_1$, the sender has an associated payload $z_{x_2}$. By OKVS correctness, $\mathsf{Decode}(P,x_2)=z_{x_2}\oplus\mathsf F^{\mathsf{igo}}(x_2)$. The receiver also obtains $x_2'=\mathsf F^{\mathsf{igo}}(x_2)$ from IGOPRF. Therefore, its output is $x_2''=z_{x_2}$. The sender likewise obtains $\mathsf F^{\mathsf{igop}}(x_1)=z_{x_1}$ for every $x_1\in X_1$.

For a query outside the all-party intersection, IGOPRF returns an independent uniform label $x_2'$. Since this label is independent of the sender's OKVS vector $P$, XORing it with $\mathsf{Decode}(P,x_2)$ preserves uniformity. Thus, the receiver learns no payload for such a query.

\begin{theorem}
In the $\mathcal{F}_{\mathsf{IGOPRF}}^t$-hybrid model with $q'=q$, the protocol $\Pi_\mathsf{IGOPPRF}^{t}$ in Figure~\ref{igoprf} realizes $\mathcal{F}_{\mathsf{IGOPPRF}}^{t}$ in Figure~\ref{figigopprf} against semi-honest adversaries corrupting any set $\mathcal C\in\mathsf{Adm}_t$.
\end{theorem}

\begin{proof}
For an element in the all-party intersection, the correctness argument above shows that the receiver obtains the same IGOPRF label as the sender. The receiver therefore recovers the associated payload by XORing this label with the corresponding OKVS value. The sender also obtains $\mathsf{F}^{\mathsf{igop}}(x_1)=\mathsf{F}^{\mathsf{igo}}(x_1)\oplus\mathsf{Decode}(P,x_1)=z_{x_1}$ for every $x_1\in X_1$. For any other receiver query, the IGOPRF output is independent and uniform. XORing it with the corresponding OKVS decoding preserves uniformity. Thus, the protocol produces the outputs specified by $\mathcal{F}_{\mathsf{IGOPPRF}}^t$.

We now consider the possible corrupted parties. Since $\mathcal C\in\mathsf{Adm}_t$, the sender and receiver cannot both be corrupted.

If the sender is corrupted, the receiver remains honest. The sender receives no additional messages after the IGOPRF invocation and computes $P$ locally from its input $\mathcal Z_1$ and the IGOPRF labels it obtains. The simulator uses the corrupted sender's input and its IGOPRF outputs to compute the same $P$. Therefore, it can reproduce the sender's view exactly.

If the receiver is corrupted, the sender remains honest. Let $r_{x_2}$ denote the IGOPRF label received by the corrupted receiver for each $x_2\in X_2$, and let $x_2''$ denote its prescribed output from $\mathcal{F}_{\mathsf{IGOPPRF}}^t$. The simulator constructs
$P^\star=\mathsf{Encode}(\{(x_2,x_2''\oplus r_{x_2})\}_{x_2\in X_2})$
and sends it to the receiver. For an intersection element, $x_2''$ is the sender's payload $z_{x_2}$ and $r_{x_2}$ is the corresponding IGOPRF label. Thus, the encoded value is exactly $z_{x_2}\oplus\mathsf{F}^{\mathsf{igo}}(x_2)$, as in the real protocol. For every other receiver query, both $x_2''$ and $r_{x_2}$ are independent uniform values, so their XOR is uniform. The real OKVS also gives pseudorandom values at these positions from the receiver's perspective. The additional sender-only keys in the real encoding likewise carry values that are hidden from the receiver by the sender's pseudorandom labels. Therefore, after accounting for the common intersection entries, OKVS obliviousness makes the real encoding and $P^\star$ computationally indistinguishable. Moreover, for every receiver query, $\mathsf{Decode}(P^\star,x_2)=x_2''\oplus r_{x_2}$, so the receiver obtains exactly its prescribed output after unmasking.

Finally, corrupted clients receive no additional messages from the IGOPPRF transformation. Their views are determined entirely by the underlying $\mathcal{F}_{\mathsf{IGOPRF}}^t$ invocation, which the simulator already reproduces. Therefore, the protocol realizes $\mathcal{F}_{\mathsf{IGOPPRF}}^t$ against every admissible semi-honest coalition.
\end{proof}

Section~\ref{extensionmpsi} uses IGOPPRF to construct several MPSI variants. Extending these constructions to the malicious setting requires an additional hashing layer that can be simulated against malicious parties, since standard cuckoo-hashing transformations do not automatically preserve malicious security~\cite{prty20}.

\section{Our MPSI}
\label{sec:mpsi}

The MPSI functionality in Figure~\ref{figfmpsi} specifies the input bounds and the effective sets used when some parties are corrupted.

\begin{figure}[htb]
\centering
\fbox{
\parbox{0.95\linewidth}{

\textbf{Parameters:} $N\geq3$ parties, corruption bound $t\in[N-1]$, and set-size bounds $n\leq n'$.

\textbf{Functionality:} For $\mathcal C\in\mathsf{Adm}_t$, set $X_i^\star=X_i$ for each honest party, where $|X_i|\leq n$. Before gate finalization, each corrupted party $\mathcal P_i$ with $i\ne1$ submits a set $X_i^\star$ of size at most $n'$, while a corrupted sender $\mathcal P_1$ submits a gate set $X_1^{\mathsf{gate}}$ of size at most $n'$. Fix all submitted sets at gate finalization. The corrupted sender then chooses $X_1^\star\subseteq X_1^{\mathsf{gate}}$ before any honest party receives its output. Once all effective sets are fixed, return $\bigcap_{i=1}^N X_i^\star$ only to $\mathcal P_2$, unless the adversary aborts before delivery.
}
}
\caption{MPSI functionality $\mathcal F_{\mathsf{MPSI}}^t$.}
\label{figfmpsi}
\end{figure}

Our MPSI protocol in Figure~\ref{mpsi} builds directly on IGOPRF. The parties first obtain gated labels for the receiver's elements and the sender's elements. The sender then pads its label list to length $n$ with independent uniform labels and sends the list in a random order. The receiver checks which of its labels appear in this list and outputs the corresponding elements.

\begin{figure}[htb]
\centering
\fbox{%
\parbox{0.95\linewidth}{%
\textbf{Inputs:} Each $\mathcal P_i$ holds a set $X_i$. Set $\mathsf{out}=\lambda+2\log_2 n$ for semi-honest security and $\mathsf{out}=\kappa+2\log_2 n$ for malicious security.
\par\smallskip
\textbf{Protocol:}
\begin{enumerate}
    \item Invoke $\mathcal F_{\mathsf{IGOPRF}}^t$ on the sets $X_i$ with $q=n$ and $q'=n'$. For each $x_2\in X_2$, $\mathcal P_2$ obtains $x_2'$.

    \item From the same invocation, $\mathcal P_1$ obtains $\mathsf F^{\mathsf{igo}}(x_1)$ for each $x_1\in X_1$.

    \item $\mathcal P_1$ pads its label list to length $n$ with independent uniform $\mathsf{out}$-bit labels and sends a random permutation $\mathcal L_1$ to $\mathcal P_2$. In the malicious protocol, $\mathcal P_2$ aborts unless $\mathcal L_1$ contains exactly $n$ labels, each of the required length.

    \item\label{step:mpsi-output} $\mathcal P_2$ outputs $\{x_2\in X_2:x_2'\in\mathcal L_1\}$.
\end{enumerate}
}}
\caption{MPSI protocol $\Pi_{\mathsf{MPSI}}$.}
\label{mpsi}
\end{figure}

In Step~\ref{step:mpsi-output}, the receiver identifies the intersection elements by matching its IGOPRF labels against the sender's permuted label list.

\begin{theorem}
In the $\mathcal{F}_{\mathsf{IGOPRF}}^t$-hybrid model with $q=n$ and $q'=n'$, $\Pi_{\mathsf{MPSI}}$ in Figure~\ref{mpsi} realizes $\mathcal{F}_{\mathsf{MPSI}}^{t}$ in Figure~\ref{figfmpsi} against malicious adversaries for every $\mathcal{C}\in\mathsf{Adm}_t$.
\end{theorem}

\begin{proof}
We first consider correctness. For every element in the all-party intersection, IGOPRF gives the receiver the same label as the sender, so the label appears in the sender's list. For an element outside the intersection, IGOPRF returns an independent uniform label to the receiver. Since the sender's list contains exactly $n$ labels, the probability that such a label matches any entry in the list is at most $n/2^{\mathsf{out}}$, which is negligible for the specified label length. Therefore, the receiver recovers exactly the intersection, except with negligible probability.

We now consider the possible corrupted parties. Since $\mathcal{C}\in\mathsf{Adm}_t$, the sender and receiver cannot both be corrupted.

If the sender is corrupted, the receiver remains honest, although some clients may also be corrupted. The simulator records the gate sets submitted by all corrupted parties and submits them at gate finalization, using $G=X_1^{\mathsf{gate}}$ for the corrupted sender. After receiving a well-formed list $\mathcal L_1$ from the sender, the simulator sets $X_1^\star=\{x\in G:\mathsf F^{\mathsf{igo}}(x)\in\mathcal L_1\}$ and submits it to the MPSI functionality. The simulator can evaluate the sender's label for every $x\in G$ without knowing the honest receiver's queries or labels. For an element belonging to every other party's effective set, the receiver obtains the same label as the sender and outputs the element exactly when its label appears in $\mathcal L_1$. This is precisely the condition captured by $X_1^\star$. An element missing from any other party's set is excluded by the ideal intersection, regardless of its sender label. Thus, the simulator correctly handles arbitrary label lists, including omissions and additional entries. Moreover, $X_1^\star\subseteq G$, so the required size bound is preserved. If the sender sends a malformed list or the IGOPRF hybrid aborts, the simulator aborts the ideal execution as well.

If the receiver is corrupted, the sender remains honest, although some clients may also be corrupted. Let $Q_2$ be the set of distinct elements in the receiver's submitted IGOPRF query batch, and let $G_2$ be its separately submitted gate set. The simulator submits $Q_2\cap G_2$ as the receiver's effective set and forwards the recorded gate sets of the other corrupted parties. Let $I$ denote the resulting ideal intersection. The simulator places the IGOPRF label received by the corrupted receiver for each $x\in I$ into the sender's list and fills the remaining $n-|I|$ positions with independent uniform labels. It then sends the list in a uniformly random order. This reproduces the honest sender's list because, for every sender element outside $I$, its label is independent of the corrupted receiver's batch outputs. The same holds for the sender's random padding labels. Repeated receiver queries reuse the same IGOPRF label. Therefore, the simulated list has the same distribution as the real list, up to negligible probability.

Finally, if only clients are corrupted, the simulator forwards their submitted gate sets to the MPSI functionality. The MPSI transformation sends these parties no additional messages, so their remaining views are already covered by the IGOPRF hybrid.

In all cases, the simulator reproduces the corrupted parties' views and the receiver's prescribed output. Therefore, $\Pi_{\mathsf{MPSI}}$ realizes $\mathcal{F}_{\mathsf{MPSI}}^t$ against every admissible malicious coalition.
\end{proof}

\subsection{Protocol Complexity}

We now analyze the total communication cost of our MPSI protocol. Let the OKVS vector length be $m=\Theta(n)$, and absorb the $\mathcal{O}(\log n)$ OKVS overhead into $\lambda$ in the semi-honest model and $\kappa$ in the malicious model. The semi-honest construction remains unchanged. Compared with the previous malicious construction, the protocol in Figure~\ref{igoprf3} introduces $N-1$ independent batch OPRF invocations, two additional OKVS encodings performed locally by $\mathcal P_2$ for the correction vectors, and $\mathcal{O}((N-1)n)$ additional local PRF evaluations and vector operations. Each non-receiver still performs only one OKVS encoding. The $N-1$ OPRF invocations can run in parallel and together incur $\mathcal{O}(Nn\kappa)$ communication. The two correction encodings add no communication.

The reconstruction of $W$ requires $\mathcal{O}(Nn\lambda)$ communication in the semi-honest model and $\mathcal{O}(Nn\kappa)$ in the malicious model. For $W'$, generating the SSM shares and reconstructing the vector through SSM/VBT costs $\mathcal{O}((N+t)n\lambda)$ communication in the semi-honest model. In the malicious model, the public masked opening within SSM and its batched verification add $\mathcal{O}(tn\kappa)$ communication using the technique of Gordon et al.~\cite{glm23}. The authenticated private-output operation also has communication linear in the vector length, bringing the total cost of this stage to $\mathcal{O}((N+t)n\kappa)$.

Finally, the sender's label list contains $n$ labels and costs $\mathcal{O}(n\lambda)$ communication in the semi-honest model or $\mathcal{O}(n\kappa)$ in the malicious model. Therefore, the total communication complexity is $\mathcal{O}((N+t)n\lambda)$ in the semi-honest model and $\mathcal{O}((N+t)n\kappa)$ in the malicious model. Both constructions require $\mathcal{O}(1)$ rounds. In particular, when $t=N-1$, the bounds become $\mathcal{O}(Nn\lambda)$ and $\mathcal{O}(Nn\kappa)$, respectively.

\section{Extensions to Other Functionalities}
\label{extensionmpsi}

IGOPPRF also supports MPSI-cardinality, threshold MPSI, and MPSI-sum. The three variants use the same hashed representation and differ only in the circuit applied to the resulting match indicators and payloads.

\begin{figure}[htb]
\centering
\fbox{
\parbox{0.95\linewidth}{
\textbf{Parameters:} $N\ge3$ parties and corruption bound $t\in[N-1]$. The threshold $\tau\in[n]$ for threshold MPSI and payload length $\ell_{\mathsf{val}}$ for MPSI-sum are public.

\textbf{Functionality:} Each $\mathcal P_i$ supplies $X_i$. Set $I=\bigcap_{i=1}^N X_i$ and deliver the output only to $\mathcal P_2$.
\par
$\mathcal F_{\mathsf{MPSICA}}^t$ returns $|I|$.
\par
$\mathcal F_{\mathsf{TMPSI}}^t$ returns $I$ if $|I|\ge\tau$ and $\bot$ otherwise.
\par
$\mathcal F_{\mathsf{MPSIsum}}^t$ takes the additional map $\mathsf{val}:X_1\to\{0,1\}^{\ell_{\mathsf{val}}}$ from $\mathcal P_1$ and returns $\sum_{x\in I}\mathsf{val}(x)$ in the application's arithmetic domain.
}
}
\caption{Ideal functionalities for the three MPSI extensions.}
\label{figfmpsica}
\label{figftmpsi}
\label{fmpsisum}
\end{figure}

The protocols use cuckoo hashing~\cite{cuckoo} and the standard $\mathcal{F}_{\mathsf{2PC}}$ functionality, which evaluates a circuit on the parties' private inputs and returns the prescribed outputs. Following Pinkas et al.~\cite{psty19}, we use a stashless cuckoo table with $b=\lceil1.27n\rceil$ bins and $k=3$ hash functions for a placement-failure target of $O(2^{-\lambda})$. IGOPPRF supplies secret-shared match indicators and payloads, and $\mathcal F_{\mathsf{2PC}}$ evaluates the variant-specific circuit.

Set $q_{\mathsf{ext}}=\max\{kn,b\}=3n$ and $\rho=\lambda+2\lceil\log_2q_{\mathsf{ext}}\rceil$. Every IGOPPRF invocation has public capacity $q_{\mathsf{ext}}$. The output length is $\rho$ for MPSI-cardinality and threshold MPSI and $\rho+\ell_{\mathsf{val}}$ for MPSI-sum. The receiver pads its $b$ table entries to $q_{\mathsf{ext}}$ with fresh dummy values and discards the corresponding outputs. For a public seed $r\in\{0,1\}^{\kappa}$, model $\mathsf H(r,\cdot)$ as a random oracle with $\kappa$-bit outputs. Equal descriptions $(x,j)$ at different parties have the same key. Except with negligible probability, distinct descriptions have distinct keys and all dummy keys are distinct from these keys and from one another.

\begin{figure}[htb]
\centering
\fbox{%
\parbox{0.95\linewidth}{%
\textbf{Inputs:} Each $\mathcal P_i$ holds $X_i$. Use $k=3$ public hash functions $h_j:\{0,1\}^*\to[b]$ for the $b=\lceil1.27n\rceil$ bins.
\par\smallskip
\textbf{Protocol:}
\begin{enumerate}
    \item\label{step:mpsica-seed} $\mathcal{P}_1$ samples $r \leftarrow \{0,1\}^\kappa$ and broadcasts it.
    \item For each $x\in X_1$ and $j\in[k]$, $\mathcal P_1$ inserts $\mathsf H(r,x\|j)$ into bin $T_S[h_j(x)]$ of its simple hash table.
    \item $\mathcal P_2$ uses cuckoo hashing to place each $x\in X_2$ in one bin $i=h_j(x)$, stores $\mathsf H(r,x\|j)$ in $T_C[i]$, and fills empty bins with independent uniform $\kappa$-bit values.
    \item Each $\mathcal{P}_i$ for $i\in[3,N]$ computes $X_i^{\mathsf{H}}=\{\mathsf{H}(r,x_i\|j)\mid x_i\in X_i,j\in[k]\}$.
    \item\label{step:mpsica-bin-labels} For each $i\in[b]$, $\mathcal{P}_1$ samples $\alpha_i\leftarrow\{0,1\}^{\rho}$.
    \item\label{step:mpsica-igopprf} In $\mathcal F_{\mathsf{IGOPPRF}}^t$ with capacity $q_{\mathsf{ext}}$, $\mathcal P_1$ programs each $u\in T_S[i]$ to $\alpha_i$, each client submits $X_i^{\mathsf H}$, and $\mathcal P_2$ queries the ordered bins of $T_C$ followed by $q_{\mathsf{ext}}-b$ fresh dummies. It keeps the first $b$ outputs $\alpha_i'$.
    \item $\mathcal P_1$ and $\mathcal P_2$ submit $(\alpha_i)_{i\in[b]}$ and $(\alpha_i')_{i\in[b]}$, respectively, to $\mathcal F_{\mathsf{2PC}}$. It sets $\mu_i=\mathbf{1}\{\alpha_i=\alpha_i'\}$ and returns $\sum_{i=1}^b\mu_i$ only to $\mathcal P_2$.
\end{enumerate}
}}
\caption{MPSI-cardinality protocol $\Pi_{\mathsf{MPSICA}}^t$.}
\label{mpsica}
\end{figure}

\begin{theorem}
In the $(\mathcal{F}_{\mathsf{IGOPPRF}}^{t},\mathcal{F}_{\mathsf{2PC}})$-hybrid model with IGOPPRF capacity $q_{\mathsf{ext}}$, $\Pi_{\mathsf{MPSICA}}^t$ in Figure~\ref{mpsica} realizes $\mathcal{F}_{\mathsf{MPSICA}}^{t}$ in Figure~\ref{figfmpsica} against semi-honest adversaries.
\end{theorem}

\begin{proof}
Condition on successful cuckoo placement and distinct keys for distinct descriptions and dummies. Each intersection element occupies one receiver bin whose key occurs in the corresponding sender bin and in every client's transformed set. IGOPPRF returns the programmed value $\alpha_i$ in each such bin and a uniform $\rho$-bit value in every other bin. With the chosen $\rho$, an accidental match in any of the $b$ comparisons has negligible probability. Thus the circuit returns $|I|$.

The only message outside $\mathcal F_{\mathsf{IGOPPRF}}^t$ and $\mathcal F_{\mathsf{2PC}}$ is the public seed $r$. Hash-table construction is local, and $\alpha_i,\alpha_i^{\prime}$ enter 2PC as private inputs. The two hybrid simulators therefore reproduce the view of every admissible semi-honest adversary from its input and prescribed output.
\end{proof}

\begin{figure}[htb]
\centering
\fbox{%
\parbox{0.95\linewidth}{%
\textbf{Inputs:} Each $\mathcal P_i$ holds $X_i$. The threshold $\tau\in[n]$ is public.
\par\smallskip
\textbf{Protocol:}
\begin{enumerate}
    \item Run Steps~\ref{step:mpsica-seed}--\ref{step:mpsica-igopprf} of $\Pi_{\mathsf{MPSICA}}^t$ in Figure~\ref{mpsica}.
    \item $\mathcal P_1$ and $\mathcal P_2$ submit $(\alpha_i)_{i\in[b]}$ and $(\alpha_i')_{i\in[b]}$, respectively, to $\mathcal F_{\mathsf{2PC}}$, which sets $\mu_i=\mathbf{1}\{\alpha_i=\alpha_i'\}$.
    \item If $\sum_{i=1}^b\mu_i\ge\tau$, the circuit returns the bits to $\mathcal P_2$, which outputs the elements in the marked bins. Otherwise, it returns $\bot$ to $\mathcal P_2$.
\end{enumerate}
}}
\caption{Threshold MPSI protocol $\Pi_{\mathsf{TMPSI}}^t$.}
\label{tmpsi}
\end{figure}

\begin{theorem}
In the $(\mathcal{F}_{\mathsf{IGOPPRF}}^{t},\mathcal{F}_{\mathsf{2PC}})$-hybrid model with IGOPPRF capacity $q_{\mathsf{ext}}$, $\Pi_{\mathsf{TMPSI}}^t$ in Figure~\ref{tmpsi} realizes $\mathcal{F}_\mathsf{TMPSI}^t$ in Figure~\ref{figftmpsi} against semi-honest adversaries.
\end{theorem}

\begin{proof}
The match bits are computed as in $\Pi_{\mathsf{MPSICA}}^t$, so the same correctness argument applies. If their sum reaches $\tau$, the circuit reveals the bits to $\mathcal P_2$, which returns the elements in the matching bins. Otherwise, $\mathcal P_2$ outputs $\bot$, as specified by $\mathcal F_{\mathsf{TMPSI}}^t$. The hybrid functionalities hide every unrevealed value.
\end{proof}

\begin{figure}[htb]
\centering
\fbox{%
\parbox{0.95\linewidth}{%
\textbf{Inputs:} Each $\mathcal P_i$ holds $X_i$, and $\mathcal P_1$ also holds the payload map $\mathsf{val}$.
\par\smallskip
\textbf{Protocol:}
\begin{enumerate}
    \item Run Steps~\ref{step:mpsica-seed}--\ref{step:mpsica-bin-labels} of $\Pi_{\mathsf{MPSICA}}^t$ in Figure~\ref{mpsica}.
    \item For each $i\in[b]$, $\mathcal{P}_1$ samples $\omega_i\leftarrow\{0,1\}^{\ell_{\mathsf{val}}}$.
    \item Run Step~\ref{step:mpsica-igopprf} of Figure~\ref{mpsica} with the following sender values. For each $x_1\in X_1$ and $j\in[k]$, $\mathcal P_1$ programs $\mathsf H(r,x_1\|j)$ to $\alpha_i\|(\omega_i\oplus\mathsf{val}(x_1))$, where $i=h_j(x_1)$. Party $\mathcal P_2$ parses each retained output as $\alpha_i'\|z_i'$.
    \item For each bin $i$, $\mathcal P_1$ submits $(\alpha_i,\omega_i)$ and $\mathcal P_2$ submits $(\alpha_i',z_i')$ to $\mathcal F_{\mathsf{2PC}}$. It sets $\mu_i=\mathbf{1}\{\alpha_i=\alpha_i'\}$ and returns $\sum_{i=1}^{b}\mu_i(z_i'\oplus\omega_i)$ only to $\mathcal P_2$.
\end{enumerate}
}}
\caption{MPSI-sum protocol $\Pi_{\mathsf{MPSIsum}}^t$.}
\label{mpsisum}
\end{figure}

\begin{theorem}
In the $(\mathcal{F}_{\mathsf{IGOPPRF}}^{t},\mathcal{F}_{\mathsf{2PC}})$-hybrid model with IGOPPRF capacity $q_{\mathsf{ext}}$, $\Pi_{\mathsf{MPSIsum}}^t$ in Figure~\ref{mpsisum} realizes $\mathcal{F}_{\mathsf{MPSIsum}}^t$ in Figure~\ref{fmpsisum} against semi-honest adversaries.
\end{theorem}

\begin{proof}
The $\rho$-bit prefix gives the same match test as in $\Pi_{\mathsf{MPSICA}}^t$. In the bin for an intersection element $x_1$, the sender programs $\alpha_i\|(\omega_i\oplus\mathsf{val}(x_1))$, so the receiver obtains $\alpha_i^{\prime}=\alpha_i$ and $z_i^{\prime}=\omega_i\oplus\mathsf{val}(x_1)$. The circuit removes $\omega_i$ and recovers $\mathsf{val}(x_1)$. Under the same conditions as in the cardinality proof, every other bin contributes zero, so the output is $\sum_{x_1\in\bigcap_{i=1}^N X_i}\mathsf{val}(x_1)$.

The mask $\omega_i$ hides the payload outside 2PC. All comparisons and additions occur inside $\mathcal F_{\mathsf{2PC}}$, and $\mathcal F_{\mathsf{IGOPPRF}}^t$ handles the remaining private values. Apart from the public seed, the two hybrid simulators reproduce the adversary's view from its prescribed input and output.
\end{proof}

\section{Experiment}
\label{expsection}

\begin{table*}[htbp]
\centering
\caption{Online running time (in seconds) and communication cost (in MB) of MPSI protocols under different input sizes $n$, numbers of parties $N$, and corruption bounds $t$. A ``$-$'' indicates that the corresponding protocol is not applicable or fails to run under the given parameter setting. In the semi-honest model, the optimal result is highlighted in \textcolor{c1}{red}, and the second-best result is highlighted in \textcolor{c2}{orange}. In the malicious model, the optimal result is highlighted in \textcolor{c3}{purple}. SH and Mal denote our semi-honest and malicious protocols, respectively.}
\resizebox{\linewidth}{!}{
\begin{tabular}{|c|c|
                c|c|
                c|c|
                c|c|
                c|c|
                c|c|
                c|c|
                c|c|}
\hline
\multirow{2}{*}{$n$} &
\multirow{2}{*}{$(N,t)$} &

\multicolumn{2}{c|}{NTY \cite{nty21}} &
\multicolumn{2}{c|}{WYC \cite{wyc24}} &
\multicolumn{2}{c|}{GLW \cite{glw24}} &
\multicolumn{2}{c|}{GGA \cite{gga25}} &
\multicolumn{2}{c|}{DCZ \cite{dcz25}} &
\multicolumn{2}{c|}{Ours (SH)} &
\multicolumn{2}{c|}{Ours (Mal)} \\

\cline{3-16}
& &
Time & Com. &
Time & Com. &
Time & Com. &
Time & Com. &
Time & Com. &
Time & Com. &
Time & Com. \\
\hline
\multicolumn{2}{|c|}{Security Model} &
\multicolumn{2}{c|}{Malicious} &
\multicolumn{2}{c|}{Semi-honest} &
\multicolumn{2}{c|}{Semi-honest} &
\multicolumn{2}{c|}{Semi-honest} &
\multicolumn{2}{c|}{Semi-honest} &
\multicolumn{2}{c|}{Semi-honest} &
\multicolumn{2}{c|}{Malicious} \\
\hline

% ====================== 2^12 ======================
\multirow{12}{*}{$2^{12}$}
   & $(4,1)$ & \cellcolor{c3}$0.166$ & \cellcolor{c3}$0.44$ & $0.050$ & \cellcolor{c1}$0.63$ & $0.346$ & $0.94$ & \cellcolor{c1}$0.001$ & \cellcolor{c2}$0.83$ & \cellcolor{c2}$0.027$ & $0.96$ & $0.409$ & $0.97$ & $1.258$ & $2.24$ \\
   & $(4,2)$ & \cellcolor{c3}$0.188$ & \cellcolor{c3}$2.39$ & $0.183$ & $3.20$ & $0.347$ & \cellcolor{c1}$0.94$ & \cellcolor{c1}$0.001$ & $1.16$ & \cellcolor{c2}$0.027$ & \cellcolor{c2}$0.96$ & $0.359$ & $1.10$ & $1.282$ & $2.63$ \\
   & $(4,3)$ & \cellcolor{c3}$0.100$ & $4.15$ & \cellcolor{c2}$0.271$ & $4.74$ & $0.302$ & \cellcolor{c1}$0.94$ & $-$ & $-$ & \cellcolor{c1}$0.029$ & \cellcolor{c2}$0.96$ & $0.344$ & $1.23$ & $1.303$ & \cellcolor{c3}$3.02$ \\
   & $(10,1)$ & \cellcolor{c3}$0.196$ & \cellcolor{c3}$1.40$ & \cellcolor{c2}$0.072$ & \cellcolor{c1}$0.86$ & $0.319$ & \cellcolor{c2}$1.44$ & \cellcolor{c1}$0.001$ & $1.82$ & $0.077$ & $2.88$ & $0.597$ & $1.75$ & $2.362$ & $4.92$ \\
   & $(10,4)$ & \cellcolor{c3}$0.255$ & $7.21$ & $0.717$ & $18.84$ & $0.358$ & \cellcolor{c1}$1.44$ & \cellcolor{c1}$0.002$ & $2.81$ & \cellcolor{c2}$0.073$ & $2.88$ & $0.656$ & \cellcolor{c2}$2.15$ & $2.288$ & \cellcolor{c3}$6.10$ \\
   & $(10,8)$ & \cellcolor{c3}$0.257$ & $14.83$ & $1.277$ & $37.34$ & $0.371$ & \cellcolor{c1}$1.44$ & \cellcolor{c1}$0.004$ & $4.13$ & \cellcolor{c2}$0.092$ & $2.88$ & $0.652$ & \cellcolor{c2}$2.67$ & $2.381$ & \cellcolor{c3}$7.67$ \\
   & $(10,9)$ & \cellcolor{c3}$0.290$ & $16.59$ & $1.320$ & $41.97$ & \cellcolor{c2}$0.344$ & \cellcolor{c1}$1.44$ & $-$ & $-$ & \cellcolor{c1}$0.081$ & $2.88$ & $0.629$ & \cellcolor{c2}$2.80$ & $2.251$ & \cellcolor{c3}$8.06$ \\
   & $(15,1)$ & \cellcolor{c3}$0.239$ & \cellcolor{c3}$2.21$ & \cellcolor{c2}$0.089$ & \cellcolor{c1}$1.05$ & $0.324$ & \cellcolor{c2}$1.85$ & \cellcolor{c1}$0.002$ & $2.64$ & $0.126$ & $4.49$ & $0.614$ & $2.41$ & $2.481$ & $7.16$ \\
   & $(15,4)$ & \cellcolor{c3}$0.263$ & \cellcolor{c3}$8.02$ & $0.975$ & $29.31$ & $0.362$ & \cellcolor{c1}$1.85$ & \cellcolor{c1}$0.002$ & $3.63$ & \cellcolor{c2}$0.114$ & $4.49$ & $0.445$ & \cellcolor{c2}$2.80$ & $1.857$ & $8.34$ \\
   & $(15,7)$ & \cellcolor{c3}$0.240$ & $13.72$ & $1.523$ & $50.90$ & $0.355$ & \cellcolor{c1}$1.85$ & \cellcolor{c1}$0.004$ & $4.62$ & \cellcolor{c2}$0.116$ & $4.49$ & $0.433$ & \cellcolor{c2}$3.20$ & $1.895$ & \cellcolor{c3}$9.52$ \\
   & $(15,13)$ & \cellcolor{c3}$0.410$ & $25.27$ & $2.420$ & $94.06$ & $0.360$ & \cellcolor{c1}$1.85$ & \cellcolor{c1}$0.008$ & $6.60$ & \cellcolor{c2}$0.132$ & $4.49$ & $0.373$ & \cellcolor{c2}$3.98$ & $1.766$ & \cellcolor{c3}$11.87$ \\
   & $(15,14)$ & \cellcolor{c3}$0.460$ & $27.05$ & $2.677$ & $101.26$ & $0.360$ & \cellcolor{c1}$1.85$ & $-$ & $-$ & \cellcolor{c1}$0.116$ & $4.49$ & \cellcolor{c2}$0.348$ & \cellcolor{c2}$4.11$ & $1.805$ & \cellcolor{c3}$12.26$ \\
\hline

% ====================== 2^16 ======================
\multirow{12}{*}{$2^{16}$}
   & $(4,1)$ & \cellcolor{c3}$0.291$ & \cellcolor{c3}$7.02$ & $0.434$ & \cellcolor{c1}$3.46$ & \cellcolor{c2}$0.414$ & \cellcolor{c2}$6.55$ & \cellcolor{c1}$0.021$ & $13.43$ & $0.544$ & $15.30$ & $0.487$ & $10.63$ & $1.527$ & $16.79$ \\
   & $(4,2)$ & \cellcolor{c3}$0.949$ & $41.19$ & $0.820$ & \cellcolor{c2}$12.20$ & \cellcolor{c2}$0.412$ & \cellcolor{c1}$6.55$ & \cellcolor{c1}$0.022$ & $18.81$ & $0.435$ & $15.30$ & $0.494$ & $12.69$ & $1.531$ & \cellcolor{c3}$22.98$ \\
   & $(4,3)$ & \cellcolor{c3}$0.793$ & $72.33$ & $0.977$ & $17.44$ & \cellcolor{c1}$0.384$ & \cellcolor{c1}$6.55$ & $-$ & $-$ & $0.509$ & $15.30$ & \cellcolor{c2}$0.496$ & \cellcolor{c2}$14.75$ & $1.575$ & \cellcolor{c3}$29.16$ \\
   & $(10,1)$ & \cellcolor{c3}$0.449$ & \cellcolor{c3}$22.07$ & $0.913$ & \cellcolor{c1}$6.87$ & \cellcolor{c2}$0.387$ & \cellcolor{c2}$14.61$ & \cellcolor{c1}$0.025$ & $29.55$ & $1.397$ & $45.92$ & $0.575$ & $23.00$ & $2.085$ & $31.25$ \\
   & $(10,4)$ & \cellcolor{c3}$1.096$ & $123.24$ & $2.314$ & $68.07$ & \cellcolor{c2}$0.410$ & \cellcolor{c1}$14.61$ & \cellcolor{c1}$0.034$ & $45.67$ & $1.369$ & $45.92$ & $0.553$ & \cellcolor{c2}$29.19$ & $2.038$ & \cellcolor{c3}$49.81$ \\
   & $(10,8)$ & \cellcolor{c3}$1.499$ & $257.63$ & $3.338$ & $131.01$ & \cellcolor{c2}$0.386$ & \cellcolor{c1}$14.61$ & \cellcolor{c1}$0.052$ & $67.16$ & $1.376$ & $45.92$ & $0.551$ & \cellcolor{c2}$37.43$ & $2.131$ & \cellcolor{c3}$74.55$ \\
   & $(10,9)$ & \cellcolor{c3}$1.480$ & $288.71$ & $3.704$ & $146.75$ & \cellcolor{c1}$0.432$ & \cellcolor{c1}$14.61$ & $-$ & $-$ & $1.369$ & $45.92$ & \cellcolor{c2}$0.585$ & \cellcolor{c2}$39.50$ & $2.090$ & \cellcolor{c3}$80.74$ \\
   & $(15,1)$ & \cellcolor{c3}$0.556$ & \cellcolor{c3}$34.63$ & $1.383$ & \cellcolor{c1}$9.72$ & \cellcolor{c2}$0.457$ & \cellcolor{c2}$21.33$ & \cellcolor{c1}$0.028$ & $42.99$ & $2.104$ & $71.42$ & $0.565$ & $33.31$ & $2.234$ & $43.30$ \\
   & $(15,4)$ & \cellcolor{c3}$1.202$ & $135.75$ & $3.123$ & $105.89$ & \cellcolor{c2}$0.413$ & \cellcolor{c1}$21.33$ & \cellcolor{c1}$0.038$ & $59.10$ & $2.147$ & $71.42$ & $0.566$ & \cellcolor{c2}$39.50$ & $2.490$ & \cellcolor{c3}$61.86$ \\
   & $(15,7)$ & \cellcolor{c3}$1.429$ & $237.19$ & $4.397$ & $179.32$ & \cellcolor{c2}$0.460$ & \cellcolor{c1}$21.33$ & \cellcolor{c1}$0.048$ & $75.22$ & $2.200$ & $71.42$ & $0.574$ & \cellcolor{c2}$45.69$ & $2.587$ & \cellcolor{c3}$80.42$ \\
   & $(15,13)$ & \cellcolor{c3}$2.120$ & $440.64$ & $6.760$ & $326.19$ & \cellcolor{c2}$0.452$ & \cellcolor{c1}$21.33$ & \cellcolor{c1}$0.100$ & $107.46$ & $2.129$ & $71.42$ & $0.597$ & \cellcolor{c2}$58.06$ & $2.507$ & \cellcolor{c3}$117.54$ \\
   & $(15,14)$ & \cellcolor{c3}$2.290$ & $470.75$ & $7.614$ & $350.67$ & \cellcolor{c1}$0.406$ & \cellcolor{c1}$21.33$ & $-$ & $-$ & $2.103$ & $71.42$ & \cellcolor{c2}$0.607$ & \cellcolor{c2}$60.12$ & $2.532$ & \cellcolor{c3}$123.72$ \\
\hline

% ====================== 2^20 ======================
\multirow{12}{*}{$2^{20}$}
   & $(4,1)$ & $8.851$ & \cellcolor{c3}$112.26$ & $10.683$ & \cellcolor{c1}$55.52$ & \cellcolor{c2}$1.536$ & \cellcolor{c2}$97.55$ & \cellcolor{c1}$1.411$ & $217.23$ & $12.494$ & $246.23$ & $2.386$ & $164.71$ & \cellcolor{c3}$4.994$ & $247.76$ \\
   & $(4,2)$ & $17.208$ & $708.66$ & $14.903$ & \cellcolor{c2}$167.99$ & \cellcolor{c2}$1.502$ & \cellcolor{c1}$97.55$ & \cellcolor{c1}$0.488$ & $304.12$ & $12.563$ & $246.23$ & $2.420$ & $197.67$ & \cellcolor{c3}$4.994$ & \cellcolor{c3}$346.65$ \\
   & $(4,3)$ & $10.510$ & $1258.34$ & $16.407$ & $235.47$ & \cellcolor{c1}$1.465$ & \cellcolor{c1}$97.55$ & $-$ & $-$ & $12.412$ & $246.23$ & \cellcolor{c2}$2.242$ & \cellcolor{c2}$230.63$ & \cellcolor{c3}$5.101$ & \cellcolor{c3}$445.54$ \\
   & $(10,1)$ & $22.433$ & \cellcolor{c3}$352.85$ & $19.372$ & \cellcolor{c1}$121.56$ & \cellcolor{c2}$2.190$ & \cellcolor{c2}$227.88$ & \cellcolor{c1}$1.750$ & $477.90$ & $38.356$ & $738.69$ & $2.997$ & $362.48$ & \cellcolor{c3}$7.841$ & $447.81$ \\
   & $(10,4)$ & $21.421$ & $2126.43$ & $32.685$ & $908.87$ & \cellcolor{c1}$2.205$ & \cellcolor{c1}$227.88$ & $8.196$ & $738.57$ & $39.143$ & $738.69$ & \cellcolor{c2}$3.035$ & \cellcolor{c2}$461.37$ & \cellcolor{c3}$8.143$ & \cellcolor{c3}$744.47$ \\
   & $(10,8)$ & $203.269$ & $4493.24$ & $46.356$ & $1718.67$ & \cellcolor{c2}$1.798$ & \cellcolor{c1}$227.88$ & \cellcolor{c1}$1.211$ & $1086.13$ & $36.503$ & $738.69$ & $3.283$ & \cellcolor{c2}$593.22$ & \cellcolor{c3}$8.827$ & \cellcolor{c3}$1140.02$ \\
   & $(10,9)$ & $-$ & $-$ & $49.752$ & $1921.13$ & \cellcolor{c1}$1.813$ & \cellcolor{c1}$227.88$ & $-$ & $-$ & $36.553$ & $738.69$ & \cellcolor{c2}$3.313$ & \cellcolor{c2}$626.18$ & \cellcolor{c3}$9.020$ & \cellcolor{c3}$1238.90$ \\
   & $(15,1)$ & $33.931$ & \cellcolor{c3}$553.35$ & $28.970$ & \cellcolor{c1}$176.59$ & \cellcolor{c2}$2.302$ & \cellcolor{c2}$336.50$ & \cellcolor{c1}$2.160$ & $695.13$ & $57.098$ & $1149.08$ & $3.619$ & $527.29$ & \cellcolor{c3}$10.551$ & $614.52$ \\
   & $(15,4)$ & $22.034$ & $2326.65$ & $49.421$ & $1413.80$ & \cellcolor{c1}$2.313$ & \cellcolor{c1}$336.50$ & $8.717$ & $955.80$ & $56.866$ & $1149.08$ & \cellcolor{c2}$3.701$ & \cellcolor{c2}$626.18$ & \cellcolor{c3}$10.900$ & \cellcolor{c3}$911.18$ \\
   & $(15,7)$ & $35.390$ & $4091.76$ & $63.742$ & $2358.57$ & \cellcolor{c1}$2.220$ & \cellcolor{c1}$336.50$ & $17.388$ & $1216.47$ & $57.067$ & $1149.08$ & \cellcolor{c2}$3.892$ & \cellcolor{c2}$725.07$ & \cellcolor{c3}$11.401$ & \cellcolor{c3}$1207.84$ \\
   & $(15,13)$ & $-$ & $-$ & $93.308$ & $4248.11$ & \cellcolor{c1}$2.654$ & \cellcolor{c1}$336.50$ & $11.511$ & $1737.81$ & $56.858$ & $1149.08$ & \cellcolor{c2}$4.237$ & \cellcolor{c2}$922.84$ & \cellcolor{c3}$12.510$ & \cellcolor{c3}$1801.16$ \\
   & $(15,14)$ & $-$ & $-$ & $96.160$ & $4563.04$ & \cellcolor{c1}$2.229$ & \cellcolor{c1}$336.50$ & $-$ & $-$ & $57.676$ & $1149.08$ & \cellcolor{c2}$5.994$ & \cellcolor{c2}$955.80$ & \cellcolor{c3}$13.128$ & \cellcolor{c3}$1900.04$ \\
\hline

\end{tabular}
}
\label{exp1}
\end{table*}

This section evaluates our MPSI protocol under LAN and WAN settings against the main implemented MPSI baselines \cite{nty21,wyc24,glw24,gga25,dcz25}. The evaluation answers three questions:
\begin{itemize}
\item \textbf{Q1}: How does our MPSI compare with state-of-the-art protocols in terms of efficiency and security guarantees?
 \item \textbf{Q2}: Does our MPSI remain competitive in WAN environments?
  \item \textbf{Q3}: Does our MPSI remain competitive under different numbers of parties and varying corruption bounds?
\end{itemize}

To answer these questions, we implemented our MPSI protocol in C++. Our full implementation is available at \url{https://github.com/ShallMate/MPSI-IGOPRF}. All experiments ran on a laptop with Ubuntu 20.04, 16 GB RAM, and an 11th Gen Intel(R) Core(TM) i7-1165G7 CPU at 2.80 GHz. Each party uses one CPU core. We set the computational security parameter to $\kappa=128$ and the statistical security parameter to $\lambda=40$. The OPRF and OKVS components use the implementations of~\cite{rs21,rr22}. The reported times cover the complete online protocol, with preprocessing generated before the timed phase. Semi-honest VBT and SSM follow~\cite{gga25,dcz25}. In the malicious experiments, the online SSM uses authenticated Beaver triples and applies the verified public opening and batched MAC check of Gordon et al.~\cite{glm23}. WAN runs use \texttt{tc} to enforce bandwidth limits of 400 Mbps and 200 Mbps with 40 ms RTT.

\noindent \textbf{Baselines}. We list below the baseline MPSI protocols used in our experiments, all of which are evaluated using the official implementations provided by their respective authors.
\begin{itemize}
  \item NTY \cite{nty21}. This is the only efficient MPSI protocol among our baselines that achieves malicious security without restricting the corrupted roles to clients, and is therefore chosen as the malicious baseline.
  \item WYC \cite{wyc24}. This is an efficient semi-honest MPSI protocol that tolerates arbitrary collusions, whose complexity scales with the corruption bound $t$, and requires no additional trust assumptions.
  \item GLW \cite{glw24}. This is a highly efficient MPSI protocol that supports functionality extensions and malicious security for clients. Like our work, it assumes that $\mathcal{P}_1$ and $\mathcal{P}_2$ do not collude. Its malicious-security guarantee covers malicious clients.
  \item GGA \cite{gga25}. This is a recently proposed efficient MPSI protocol in the semi-honest model. At $t = N-1$, it requires a general MPC protocol or a trusted third party.
  \item DCZ \cite{dcz25}. This is a recently proposed semi-honest MPSI protocol that supports functionality extensions. Its complexity is independent of $t$.
\end{itemize}

\subsection{Evaluation for Q1}

We evaluate all protocols with input sizes $n \in \{2^{12},2^{16},2^{20}\}$, numbers of parties $N \in \{4,10,15\}$, and varying corruption bounds $t$. Table~\ref{exp1} reports the online runtime and communication of the baselines and our two variants.

\noindent\textbf{NTY}~\cite{nty21}. The baseline-to-ours runtime ratio spans $0.08\times$--$23.0\times$. At $n=2^{20}$ and $(N,t)=(10,8)$, our malicious protocol reduces runtime from $203.269$ to $8.827$ seconds. The communication ratio spans $0.20\times$--$3.94\times$, and the same setting reduces communication from $4493.24$ to $1140.02$~MB. The gains increase with the input size and corruption bound.

\noindent\textbf{WYC}~\cite{wyc24}. The runtime ratio spans $0.12\times$--$22.0\times$. At $n=2^{20}$ and $(N,t)=(15,13)$, our semi-honest protocol completes in $4.237$ seconds, compared with $93.308$ seconds for WYC. The communication ratio spans $0.29\times$--$24.7\times$. WYC's dependence on $t$ produces the largest communication gaps at high corruption bounds.

\noindent\textbf{GLW}~\cite{glw24}. GLW gives lower raw semi-honest cost in most evaluated settings. Its runtime and communication ratios span $0.37\times$--$1.03\times$ and $0.35\times$--$0.97\times$, respectively. At $n=2^{12}$ and $(N,t)=(15,14)$, the runtimes are close: $0.360$ seconds for GLW and $0.348$ seconds for ours. Under the same non-collusion assumption, our malicious protocol covers every admissible corrupted role, whereas GLW's malicious extension covers clients.

\noindent\textbf{GGA}~\cite{gga25}. GGA gives the lowest runtime on many small inputs. The runtime ratio spans $0.002\times$--$4.47\times$, and the communication ratio spans $0.85\times$--$1.88\times$. At $n=2^{20}$ and $(N,t)=(15,7)$, our protocol takes $3.892$ seconds and $725.07$~MB, compared with $17.388$ seconds and $1216.47$~MB for GGA.

\noindent\textbf{DCZ}~\cite{dcz25}. The runtime and communication ratios span $0.07\times$--$15.8\times$ and $0.78\times$--$2.18\times$, respectively. At $n=2^{20}$ and $(N,t)=(15,1)$, our protocol reduces runtime from $57.098$ to $3.619$ seconds and communication from $1149.08$ to $527.29$~MB. The $t$-adaptive construction produces its largest communication gain at small $t$.

These results answer \textbf{Q1}. Large inputs expose the main runtime gains, and threshold adaptivity reduces communication as $t$ decreases. The malicious protocol combines these gains with security for every admissible corrupted role under the designated sender-receiver non-collusion assumption.

\begin{table*}[htbp]
\centering
\caption{Online runtime (in seconds) of MPSI protocols in the WAN setting under bandwidths of 400 Mbps and 200 Mbps, with varying input size $n$, number of parties $N$, and corruption bound $t$. A ``$-$'' indicates that the protocol is inapplicable or fails under the corresponding setting. In the semi-honest model, the best and second-best results are highlighted in \textcolor{c1}{red} and \textcolor{c2}{orange}, respectively. In the malicious model, the best result is highlighted in \textcolor{c3}{purple}. SH and Mal denote our semi-honest and malicious protocols, respectively.}
\resizebox{\linewidth}{!}{
\begin{tabular}{|c|c|
                c|c|c|c|c|c|c|
                c|c|c|c|c|c|c|}
\hline
\multirow{3}{*}{$n$} &
\multirow{3}{*}{$(N,t)$} &

\multicolumn{7}{c|}{$400$ Mbps} &
\multicolumn{7}{c|}{$200$ Mbps} \\
\cline{3-16}

& &
NTY \cite{nty21} & WYC \cite{wyc24} & GLW \cite{glw24} & GGA \cite{gga25} & DCZ \cite{dcz25} & Ours (SH) & Ours (Mal) &
NTY \cite{nty21} & WYC \cite{wyc24} & GLW \cite{glw24} & GGA \cite{gga25} & DCZ \cite{dcz25} & Ours (SH) & Ours (Mal) \\
\cline{3-16}

& &
Time & Time & Time & Time & Time & Time & Time &
Time & Time & Time & Time & Time & Time & Time \\
\hline

\multicolumn{2}{|c|}{Security Model} &
Malicious & Semi-honest & Semi-honest & Semi-honest & Semi-honest & Semi-honest & Malicious &
Malicious & Semi-honest & Semi-honest & Semi-honest & Semi-honest & Semi-honest & Malicious \\
\hline

\multirow{12}{*}{$2^{12}$}
  & $(4,1)$       & \cellcolor{c3}$0.17$ & $0.06$ & $0.32$ & \cellcolor{c1}$0.01$ & \cellcolor{c2}$0.05$ & $1.16$ & $3.13$
           & \cellcolor{c3}$0.16$ & $0.07$ & $0.34$ & \cellcolor{c1}$0.02$ & \cellcolor{c2}$0.07$ & $1.16$ & $3.18$ \\
  & $(4,2)$       & \cellcolor{c3}$0.23$ & $0.26$ & $0.32$ & \cellcolor{c1}$0.01$ & \cellcolor{c2}$0.05$ & $1.17$ & $3.21$
           & \cellcolor{c3}$0.29$ & $0.33$ & $0.34$ & \cellcolor{c1}$0.02$ & \cellcolor{c2}$0.07$ & $1.17$ & $3.22$ \\
  & $(4,3)$       & \cellcolor{c3}$0.16$ & $0.37$ & \cellcolor{c2}$0.34$ & $-$ & \cellcolor{c1}$0.05$ & $1.17$ & $3.31$
           & \cellcolor{c3}$0.26$ & $0.51$ & \cellcolor{c2}$0.37$ & $-$ & \cellcolor{c1}$0.07$ & $1.18$ & $3.76$ \\
  & $(10,1)$       & \cellcolor{c3}$0.22$ & \cellcolor{c2}$0.08$ & $0.34$ & \cellcolor{c1}$0.02$ & $0.13$ & $1.45$ & $4.34$
           & \cellcolor{c3}$0.24$ & \cellcolor{c2}$0.10$ & $0.37$ & \cellcolor{c1}$0.04$ & $0.19$ & $1.30$ & $4.19$ \\
  & $(10,4)$       & \cellcolor{c3}$0.33$ & $0.95$ & $0.33$ & \cellcolor{c1}$0.03$ & \cellcolor{c2}$0.13$ & $1.48$ & $4.55$
           & \cellcolor{c3}$0.50$ & $1.45$ & $0.37$ & \cellcolor{c1}$0.06$ & \cellcolor{c2}$0.19$ & $1.44$ & $4.42$ \\
  & $(10,8)$       & \cellcolor{c3}$0.52$ & $1.67$ & $0.34$ & \cellcolor{c1}$0.05$ & \cellcolor{c2}$0.15$ & $1.83$ & $4.45$
           & \cellcolor{c3}$0.83$ & $2.42$ & $0.37$ & \cellcolor{c1}$0.09$ & \cellcolor{c2}$0.21$ & $1.64$ & $4.54$ \\
  & $(10,9)$       & \cellcolor{c3}$0.61$ & $1.87$ & \cellcolor{c2}$0.34$ & $-$ & \cellcolor{c1}$0.14$ & $1.76$ & $4.63$
           & \cellcolor{c3}$0.94$ & $2.66$ & \cellcolor{c2}$0.37$ & $-$ & \cellcolor{c1}$0.20$ & $1.78$ & $4.45$ \\
  & $(15,1)$       & \cellcolor{c3}$0.27$ & \cellcolor{c2}$0.10$ & $0.35$ & \cellcolor{c1}$0.03$ & $0.22$ & $1.45$ & $4.44$
           & \cellcolor{c3}$0.32$ & \cellcolor{c2}$0.11$ & $0.41$ & \cellcolor{c1}$0.05$ & $0.31$ & $1.39$ & $3.91$ \\
  & $(15,4)$       & \cellcolor{c3}$0.36$ & $1.47$ & $0.36$ & \cellcolor{c1}$0.04$ & \cellcolor{c2}$0.20$ & $1.29$ & $3.74$
           & \cellcolor{c3}$0.52$ & $2.52$ & $0.40$ & \cellcolor{c1}$0.07$ & \cellcolor{c2}$0.29$ & $1.32$ & $4.07$ \\
  & $(15,7)$       & \cellcolor{c3}$0.46$ & $2.38$ & $0.39$ & \cellcolor{c1}$0.05$ & \cellcolor{c2}$0.21$ & $1.41$ & $3.77$
           & \cellcolor{c3}$0.78$ & $3.12$ & $0.40$ & \cellcolor{c1}$0.10$ & \cellcolor{c2}$0.30$ & $1.49$ & $3.90$ \\
  & $(15,13)$      & \cellcolor{c3}$0.90$ & $3.46$ & $0.36$ & \cellcolor{c1}$0.07$ & \cellcolor{c2}$0.22$ & $1.76$ & $3.81$
           & \cellcolor{c3}$1.40$ & $5.54$ & $0.41$ & \cellcolor{c1}$0.14$ & \cellcolor{c2}$0.31$ & $1.61$ & $3.92$ \\
  & $(15,14)$      & \cellcolor{c3}$0.98$ & $3.73$ & \cellcolor{c2}$0.38$ & $-$ & \cellcolor{c1}$0.21$ & $1.56$ & $3.72$
           & \cellcolor{c3}$1.52$ & $5.54$ & \cellcolor{c2}$0.41$ & $-$ & \cellcolor{c1}$0.30$ & $1.71$ & $3.85$ \\
\hline

\multirow{12}{*}{$2^{16}$}
  & $(4,1)$       & \cellcolor{c3}$0.45$ & $0.55$ & \cellcolor{c2}$0.52$ & \cellcolor{c1}$0.16$ & $0.85$ & $2.44$ & $3.81$
           & \cellcolor{c3}$0.57$ & \cellcolor{c2}$0.60$ & $0.68$ & \cellcolor{c1}$0.29$ & $1.16$ & $2.59$ & $4.18$ \\
  & $(4,2)$       & \cellcolor{c3}$1.39$ & $0.99$ & \cellcolor{c2}$0.51$ & \cellcolor{c1}$0.21$ & $0.74$ & $2.38$ & $4.47$
           & \cellcolor{c3}$2.21$ & $1.21$ & \cellcolor{c2}$0.64$ & \cellcolor{c1}$0.40$ & $1.05$ & $2.53$ & $4.78$ \\
  & $(4,3)$       & \cellcolor{c3}$1.79$ & $1.25$ & \cellcolor{c1}$0.51$ & $-$ & \cellcolor{c2}$0.82$ & $2.45$ & $4.57$
           & \cellcolor{c3}$3.25$ & $1.64$ & \cellcolor{c1}$0.69$ & $-$ & \cellcolor{c2}$1.12$ & $2.69$ & $5.10$ \\
  & $(10,1)$       & \cellcolor{c3}$0.90$ & $1.10$ & \cellcolor{c2}$0.72$ & \cellcolor{c1}$0.32$ & $2.32$ & $5.56$ & $4.45$
           & \cellcolor{c3}$1.36$ & $1.34$ & \cellcolor{c2}$1.07$ & \cellcolor{c1}$0.62$ & $3.23$ & $5.91$ & $4.87$ \\
  & $(10,4)$       & \cellcolor{c3}$3.06$ & $2.91$ & \cellcolor{c2}$0.71$ & \cellcolor{c1}$0.49$ & $2.29$ & $5.19$ & $5.55$
           & \cellcolor{c3}$5.64$ & $4.37$ & \cellcolor{c2}$1.04$ & \cellcolor{c1}$0.95$ & $3.21$ & $5.72$ & $6.53$ \\
  & $(10,8)$       & \cellcolor{c3}$5.89$ & $4.55$ & \cellcolor{c2}$0.74$ & \cellcolor{c1}$0.72$ & $2.29$ & $5.11$ & $6.78$
           & $11.30$ & $7.13$ & \cellcolor{c1}$0.99$ & \cellcolor{c2}$1.40$ & $3.21$ & $5.63$ & \cellcolor{c3}$8.06$ \\
  & $(10,9)$       & \cellcolor{c3}$7.02$ & $4.95$ & \cellcolor{c1}$0.70$ & $-$ & \cellcolor{c2}$2.29$ & $5.21$ & $7.06$
           & $12.80$ & $7.98$ & \cellcolor{c1}$1.06$ & $-$ & \cellcolor{c2}$3.21$ & $5.78$ & \cellcolor{c3}$8.24$ \\
  & $(15,1)$       & \cellcolor{c3}$1.30$ & $1.67$ & \cellcolor{c2}$0.87$ & \cellcolor{c1}$0.46$ & $3.53$ & $7.72$ & $4.84$
           & \cellcolor{c3}$2.02$ & $1.91$ & \cellcolor{c2}$1.30$ & \cellcolor{c1}$0.89$ & $4.96$ & $8.52$ & $5.64$ \\
  & $(15,4)$       & \cellcolor{c3}$3.38$ & $4.18$ & \cellcolor{c2}$0.84$ & \cellcolor{c1}$0.63$ & $3.58$ & $7.60$ & $6.10$
           & \cellcolor{c3}$6.21$ & $6.32$ & \cellcolor{c2}$1.35$ & \cellcolor{c1}$1.22$ & $5.00$ & $8.35$ & $7.03$ \\
  & $(15,7)$       & \cellcolor{c3}$5.45$ & $6.05$ & \cellcolor{c2}$0.85$ & \cellcolor{c1}$0.80$ & $3.63$ & $7.60$ & $7.00$
           & $10.44$ & $9.77$ & \cellcolor{c1}$1.34$ & \cellcolor{c2}$1.55$ & $5.06$ & $8.35$ & \cellcolor{c3}$8.35$ \\
  & $(15,13)$      & $10.58$ & $9.91$ & \cellcolor{c1}$0.83$ & \cellcolor{c2}$1.17$ & $3.56$ & $7.30$ & \cellcolor{c3}$8.74$
           & $19.39$ & $16.00$ & \cellcolor{c1}$1.35$ & \cellcolor{c2}$2.25$ & $4.99$ & $8.24$ & \cellcolor{c3}$10.85$ \\
  & $(15,14)$      & $11.33$ & $10.40$ & \cellcolor{c1}$0.84$ & $-$ & \cellcolor{c2}$3.53$ & $7.46$ & \cellcolor{c3}$8.99$
           & $20.74$ & $17.23$ & \cellcolor{c1}$1.30$ & $-$ & \cellcolor{c2}$4.96$ & $8.41$ & \cellcolor{c3}$11.20$ \\
\hline

\multirow{12}{*}{$2^{20}$}
  & $(4,1)$       & \cellcolor{c3}$10.97$ & $11.74$ & \cellcolor{c1}$3.46$ & \cellcolor{c2}$3.58$ & $17.42$ & $8.07$ & $13.18$
           & \cellcolor{c3}$13.19$ & $12.45$ & \cellcolor{c1}$5.64$ & \cellcolor{c2}$5.76$ & $22.34$ & $11.33$ & $18.38$ \\
  & $(4,2)$       & $29.22$ & $18.05$ & \cellcolor{c1}$3.48$ & \cellcolor{c2}$3.53$ & $17.49$ & $9.19$ & \cellcolor{c3}$16.18$
           & $43.26$ & $21.24$ & \cellcolor{c1}$5.44$ & \cellcolor{c2}$6.57$ & $22.41$ & $13.12$ & \cellcolor{c3}$23.03$ \\
  & $(4,3)$       & $31.02$ & $20.45$ & \cellcolor{c1}$3.49$ & $-$ & $17.34$ & \cellcolor{c2}$10.17$ & \cellcolor{c3}$19.36$
           & $56.91$ & $25.48$ & \cellcolor{c1}$5.43$ & $-$ & $22.26$ & \cellcolor{c2}$14.74$ & \cellcolor{c3}$28.02$ \\
  & $(10,1)$       & $29.70$ & $21.48$ & \cellcolor{c2}$6.84$ & \cellcolor{c1}$6.53$ & $53.13$ & $15.80$ & \cellcolor{c3}$20.02$
           & $37.15$ & $23.25$ & \cellcolor{c2}$11.62$ & \cellcolor{c1}$11.31$ & $67.90$ & $23.10$ & \cellcolor{c3}$28.96$ \\
  & $(10,4)$       & $56.83$ & $44.27$ & \cellcolor{c1}$7.07$ & \cellcolor{c2}$15.58$ & $53.92$ & $19.13$ & \cellcolor{c3}$29.65$
           & $101.80$ & $62.12$ & \cellcolor{c1}$11.25$ & \cellcolor{c2}$22.97$ & $68.69$ & $28.34$ & \cellcolor{c3}$44.33$ \\
  & $(10,8)$       & $136.85$ & $71.94$ & \cellcolor{c1}$6.47$ & \cellcolor{c2}$12.07$ & $51.28$ & $23.86$ & \cellcolor{c3}$42.77$
           & $423.06$ & $101.14$ & \cellcolor{c1}$11.26$ & \cellcolor{c2}$22.93$ & $66.05$ & $35.72$ & \cellcolor{c3}$65.65$ \\
  & $(10,9)$       & $-$ & $70.86$ & \cellcolor{c1}$6.45$ & $-$ & $51.33$ & \cellcolor{c2}$25.01$ & \cellcolor{c3}$46.45$
           & $-$ & $109.94$ & \cellcolor{c1}$11.22$ & $-$ & $66.10$ & \cellcolor{c2}$37.25$ & \cellcolor{c3}$70.59$ \\
  & $(15,1)$       & $45.22$ & $32.00$ & \cellcolor{c2}$9.16$ & \cellcolor{c1}$9.11$ & $80.08$ & $22.20$ & \cellcolor{c3}$25.73$
           & $56.97$ & $35.88$ & \cellcolor{c2}$16.16$ & \cellcolor{c1}$16.06$ & $103.06$ & $32.76$ & \cellcolor{c3}$37.68$ \\
  & $(15,4)$       & $61.43$ & $66.28$ & \cellcolor{c1}$9.00$ & \cellcolor{c2}$18.27$ & $79.85$ & $25.51$ & \cellcolor{c3}$34.93$
           & $110.50$ & $91.64$ & \cellcolor{c1}$16.23$ & \cellcolor{c2}$27.83$ & $102.83$ & $37.90$ & \cellcolor{c3}$53.43$ \\
  & $(15,7)$       & $133.34$ & $88.44$ & \cellcolor{c1}$9.19$ & $29.55$ & $80.05$ & \cellcolor{c2}$29.04$ & \cellcolor{c3}$44.79$
           & $208.56$ & $136.08$ & \cellcolor{c1}$16.51$ & \cellcolor{c2}$41.72$ & $103.03$ & $43.39$ & \cellcolor{c3}$69.70$ \\
  & $(15,13)$      & $-$ & $135.28$ & \cellcolor{c1}$9.29$ & \cellcolor{c2}$28.89$ & $79.84$ & $36.23$ & \cellcolor{c3}$66.00$
           & $-$ & $215.65$ & \cellcolor{c1}$16.29$ & \cellcolor{c2}$46.27$ & $102.82$ & $54.22$ & \cellcolor{c3}$102.05$ \\
  & $(15,14)$      & $-$ & $142.27$ & \cellcolor{c1}$9.26$ & $-$ & $80.66$ & \cellcolor{c2}$37.80$ & \cellcolor{c3}$68.97$
           & $-$ & $230.05$ & \cellcolor{c1}$16.60$ & $-$ & $103.64$ & \cellcolor{c2}$56.17$ & \cellcolor{c3}$114.15$ \\
\hline

\end{tabular}
}
\label{exp2}
\end{table*}

\subsection{Evaluation for Q2}

We evaluate our MPSI in WAN settings with bandwidths of $400$~Mbps and $200$~Mbps, and compare it with existing MPSI protocols~\cite{nty21,wyc24,glw24,gga25,dcz25}. Since communication cost is independent of bandwidth, we focus on runtime under bandwidth constraints. The results are summarized in Table~\ref{exp2}.

\noindent\textbf{NTY}~\cite{nty21}. We compare NTY with our malicious protocol. The runtime ratio spans $0.05\times$--$3.20\times$ at $400$~Mbps and $0.05\times$--$6.44\times$ at $200$~Mbps. At $n=2^{20}$ and $(N,t)=(10,8)$, NTY takes $136.85$ and $423.06$ seconds, whereas ours takes $42.77$ and $65.65$ seconds, respectively.

\noindent\textbf{WYC}~\cite{wyc24}. The runtime ratio spans $0.05\times$--$3.76\times$ at $400$~Mbps and $0.06\times$--$4.10\times$ at $200$~Mbps. At $n=2^{20}$ and $(N,t)=(15,14)$, WYC takes $142.27$ and $230.05$ seconds, whereas ours takes $37.80$ and $56.17$ seconds.

\noindent\textbf{GLW}~\cite{glw24}. GLW gives lower semi-honest WAN runtime across the evaluated points. The runtime ratio spans $0.11\times$--$0.43\times$ at $400$~Mbps and $0.15\times$--$0.50\times$ at $200$~Mbps. Its malicious extension covers clients, and our malicious protocol covers every admissible corrupted role under the same non-collusion assumption.

\noindent\textbf{GGA}~\cite{gga25}. The runtime ratio spans $0.008\times$--$1.02\times$ at $400$~Mbps and $0.015\times$--$0.96\times$ at $200$~Mbps. At $n=2^{20}$ and $(N,t)=(15,7)$, our protocol takes $29.04$ seconds at $400$~Mbps, compared with $29.55$ seconds for GGA. At $200$~Mbps, the corresponding times are $43.39$ and $41.72$ seconds.

\noindent\textbf{DCZ}~\cite{dcz25}. The runtime ratio spans $0.04\times$--$3.61\times$ at $400$~Mbps and $0.06\times$--$3.15\times$ at $200$~Mbps. At $n=2^{20}$ and $(N,t)=(15,1)$, DCZ takes $80.08$ and $103.06$ seconds, whereas ours takes $22.20$ and $32.76$ seconds. Threshold adaptivity retains its communication-derived benefit under both bandwidth limits.

These results answer \textbf{Q2}. The large-input gains over NTY, WYC, and DCZ persist in WAN, and the malicious protocol reaches a $6.44\times$ speedup over NTY at $200$~Mbps.

\subsection{Evaluation for Q3}

\begin{figure}[htbp] % 两栏论文可改为 figure*
  \centering

  % ---- Row 1 ----
  \begin{subfigure}[t]{0.22\textwidth}
    \centering
    \includegraphics[width=\linewidth]{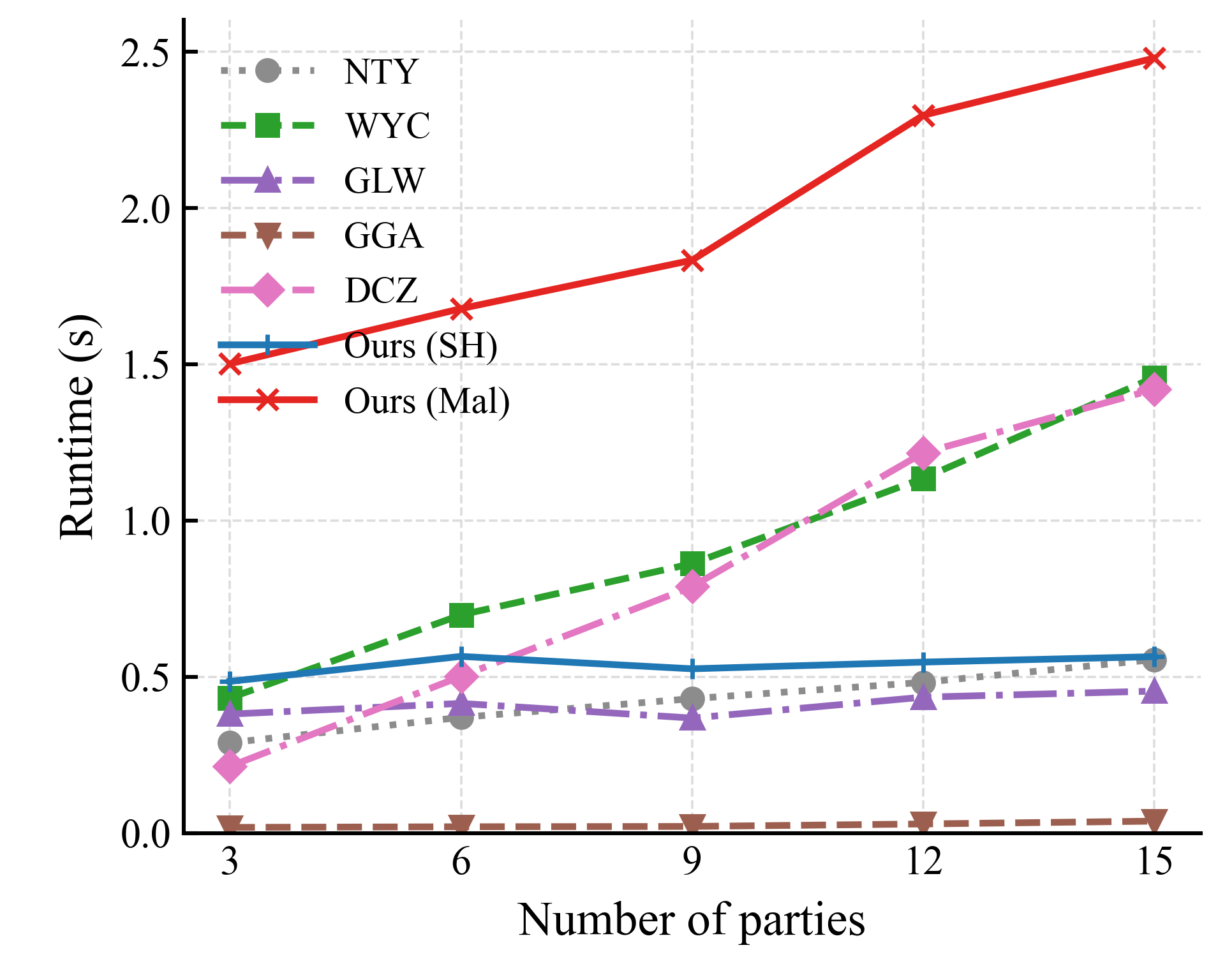}
    \caption{Runtime vs. number of parties $N$ ($t = 1$, $n = 2^{16}$).}
    \label{p1}
  \end{subfigure}\hfill
  \begin{subfigure}[t]{0.22\textwidth}
    \centering
    \includegraphics[width=\linewidth]{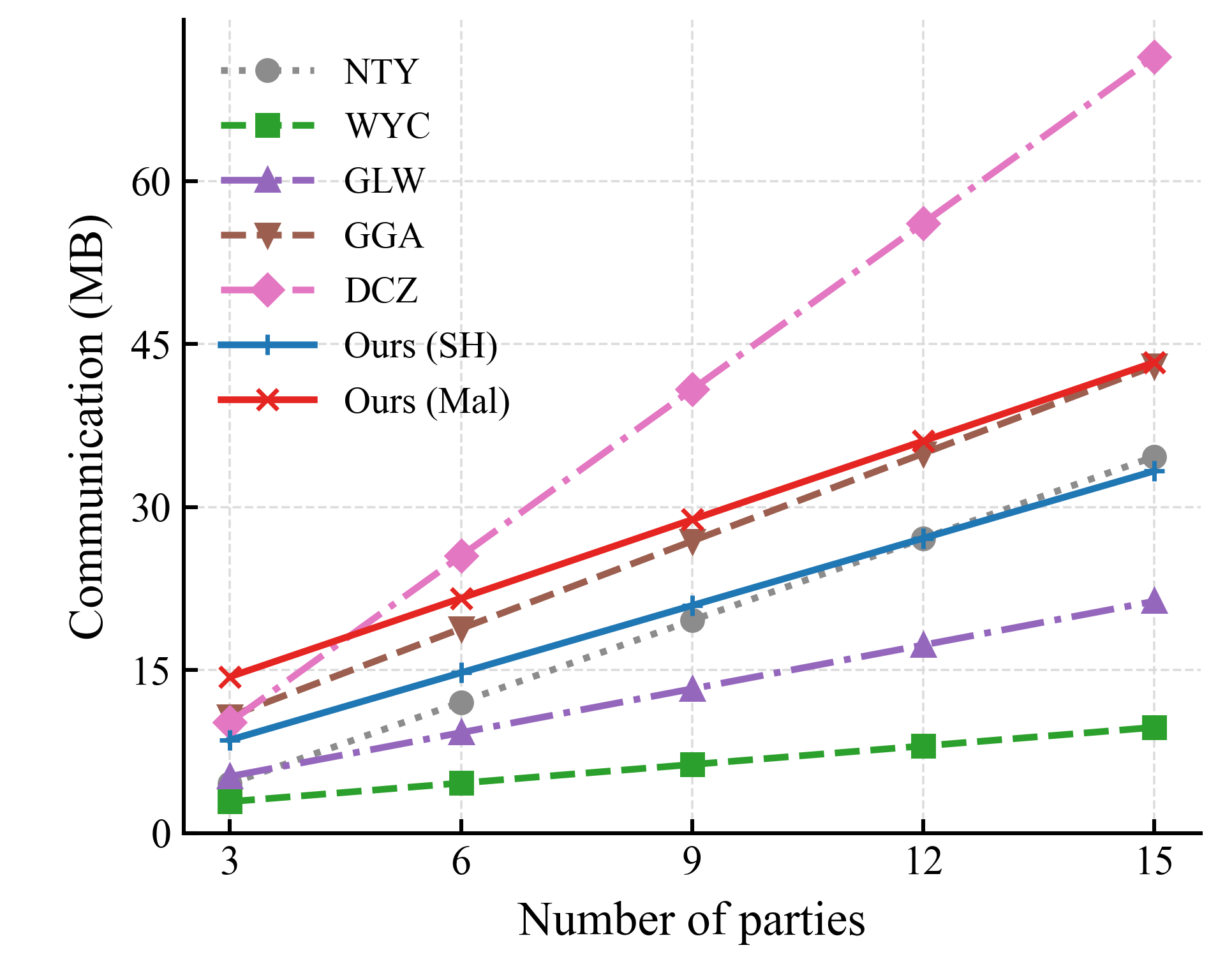}
    \caption{Communication vs. number of parties $N$ ($t = 1$, $n = 2^{16}$).}
    \label{p2}
  \end{subfigure}

  % ---- Row 2 ----
  \begin{subfigure}[t]{0.22\textwidth}
    \centering
    \includegraphics[width=\linewidth]{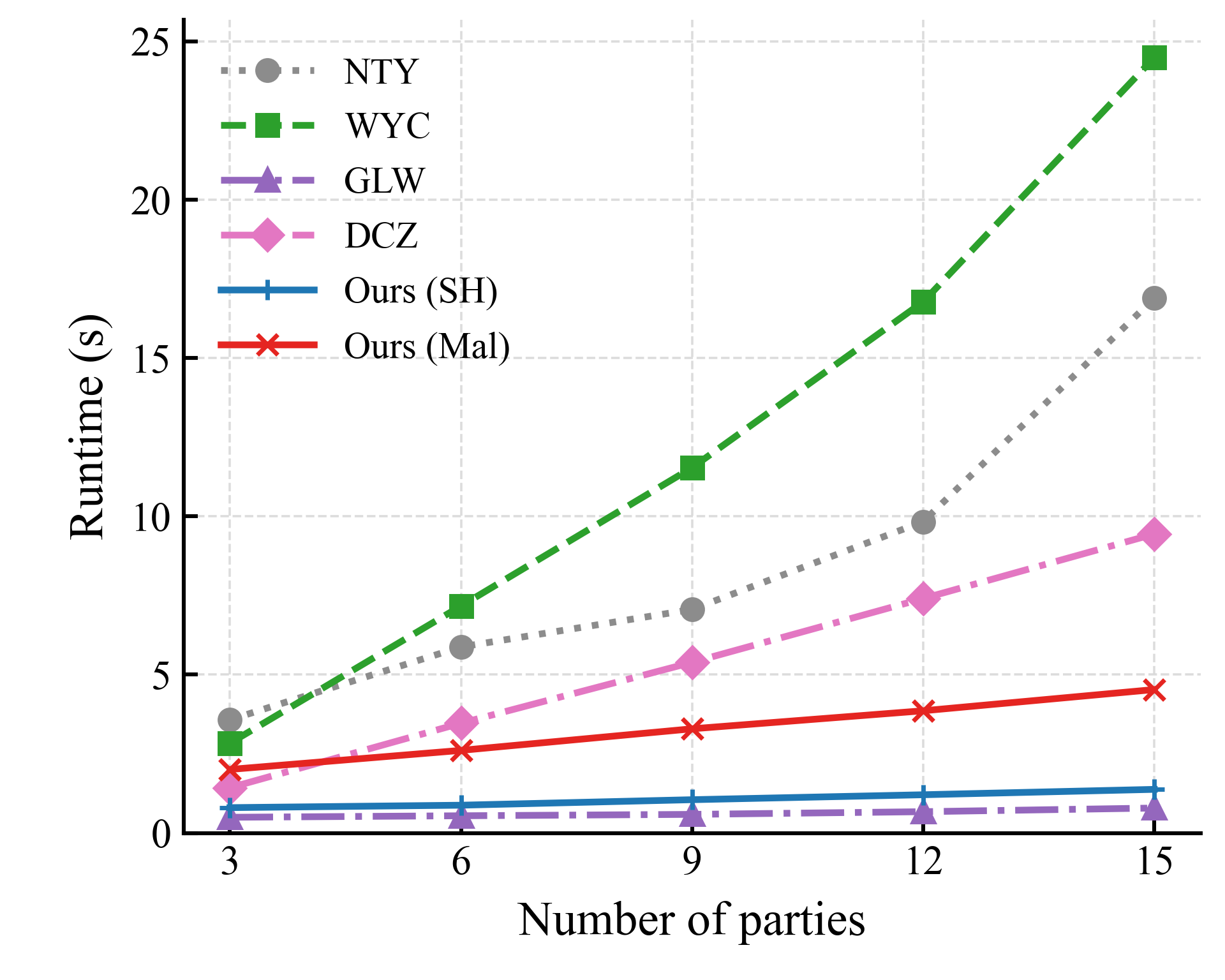}
    \caption{Runtime vs. number of parties $N$ ($t = N-1$, $n = 2^{18}$).}
    \label{p3}
  \end{subfigure}\hfill
  \begin{subfigure}[t]{0.22\textwidth}
    \centering
    \includegraphics[width=\linewidth]{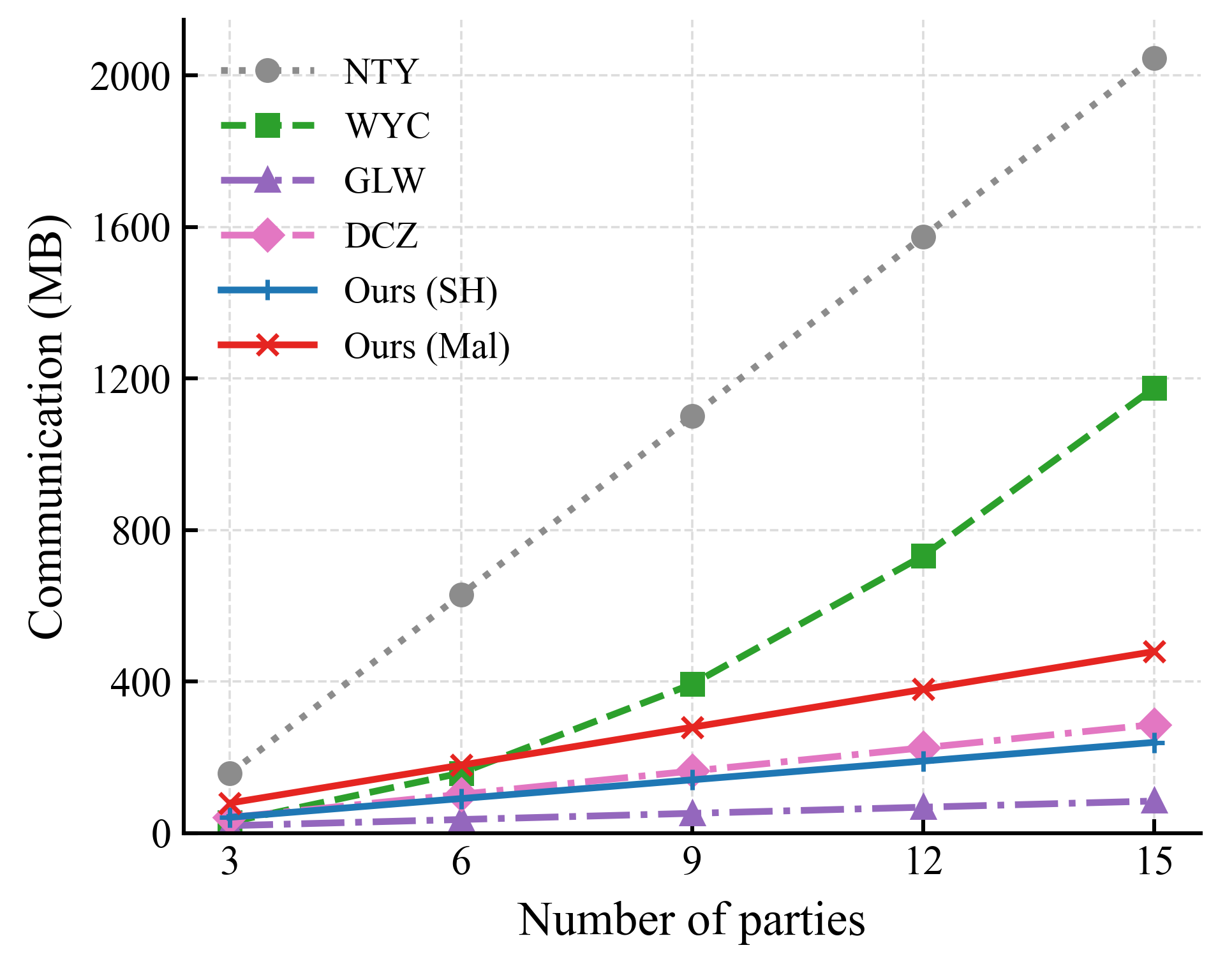}
    \caption{Communication vs. number of parties $N$ ($t = N-1$, $n = 2^{18}$).}
    \label{p4}
  \end{subfigure}

  % ---- Row 3 ----
  \begin{subfigure}[t]{0.22\textwidth}
    \centering
    \includegraphics[width=\linewidth]{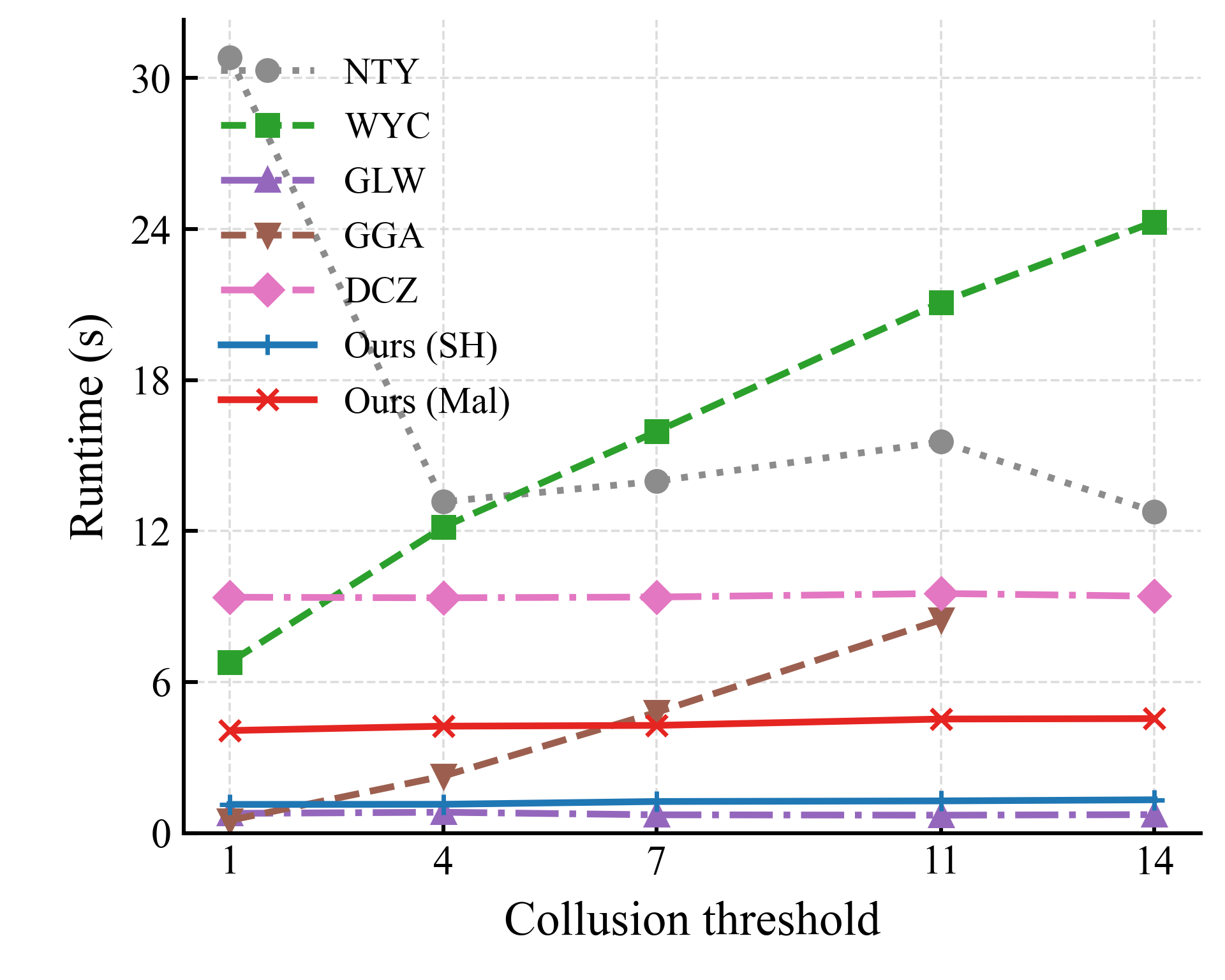}
    \caption{Runtime vs. corruption bound $t$ ($N = 15$, $n = 2^{18}$).}
    \label{p5}
  \end{subfigure}\hfill
  \begin{subfigure}[t]{0.22\textwidth}
    \centering
    \includegraphics[width=\linewidth]{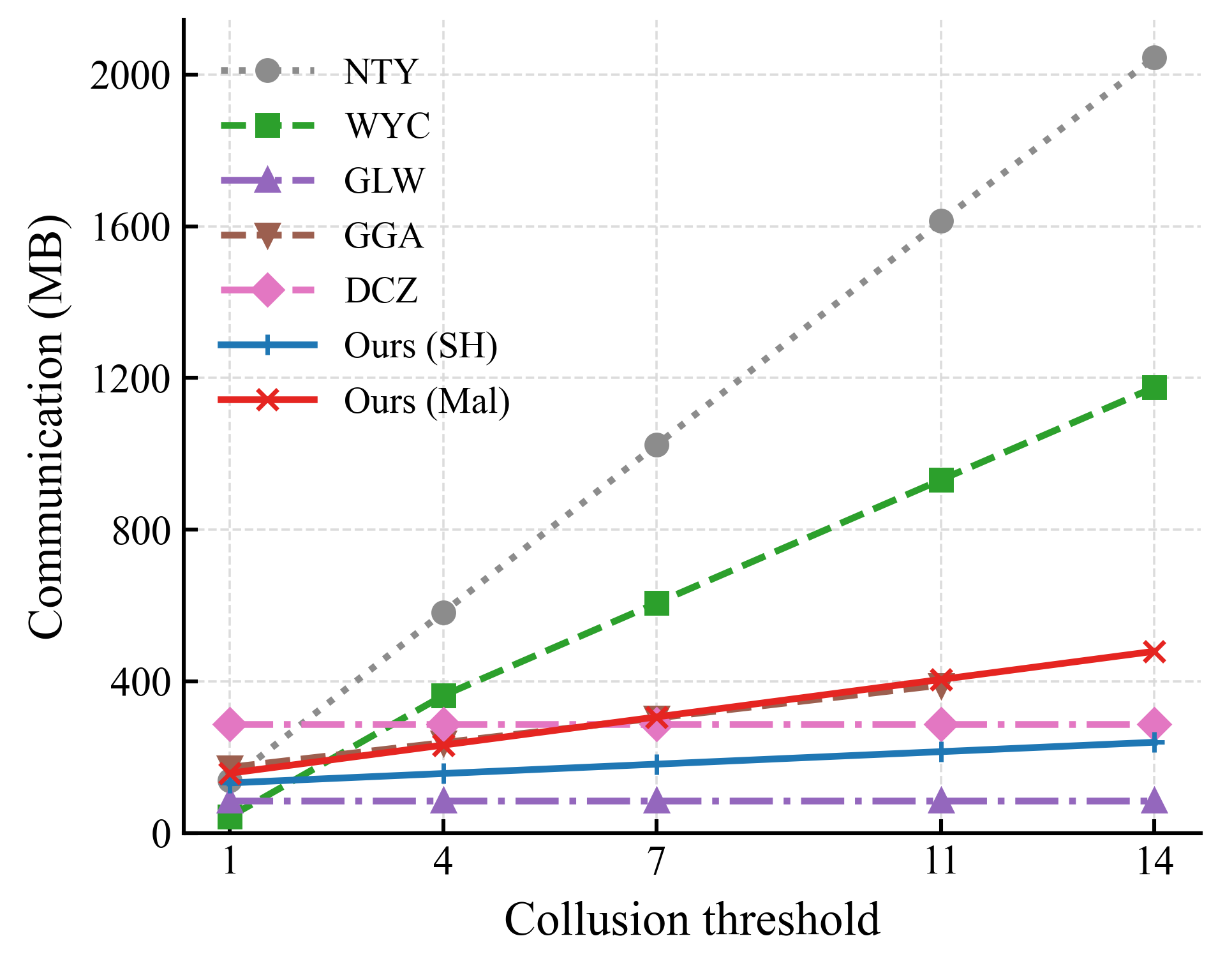}
    \caption{\! Communication vs. corruption bound $t$ ($N\!=\!15, n = 2^{18}$).}
    \label{p6}
  \end{subfigure}

  \caption{Runtime and communication vs. number of parties $N$ and corruption bound $t$. GGA has no point at $t=N-1$, where its construction requires a generic MPC protocol or a trusted party.}
  \label{exp3}
\end{figure}

Figure~\ref{exp3} answers \textbf{Q3} by varying the number of parties and the corruption bound.

We first fix $n=2^{16}$ and $t=1$. As $N$ increases from $3$ to $15$, the semi-honest runtime remains between $0.49$ and $0.56$ seconds, and the malicious runtime increases from $1.50$ to $2.48$ seconds. Communication grows linearly from $8.56$ to $33.31$~MB in the semi-honest model and from $14.38$ to $43.30$~MB in the malicious model. GGA gives the lowest semi-honest runtime, and NTY gives the lowest malicious runtime at this input size.

We next set $n=2^{18}$ and $t=N-1$. The semi-honest runtime grows from $0.80$ to $1.38$ seconds and remains second to GLW. The malicious runtime grows from $2.01$ to $4.53$ seconds, giving a $1.8\times$--$3.7\times$ speedup over NTY. Its communication is $2.0\times$--$4.3\times$ lower than NTY. From $N=6$ onward, the semi-honest protocol communicates less than WYC and DCZ, with GLW retaining the lowest communication.

Finally, we fix $N=15$ and $n=2^{18}$ and vary $t$. The semi-honest runtime changes from $1.14$ to $1.32$ seconds, and the malicious runtime changes from $4.07$ to $4.54$ seconds. Communication grows from $132.12$ to $239.26$~MB and from $157.69$ to $479.13$~MB, respectively. The malicious protocol is faster than NTY at every evaluated $t$ and communicates less for $t\geq4$.

Q3 therefore shows near-linear communication growth in $N$, modest runtime sensitivity to $t$, and the strongest malicious gains at large corruption bounds.

\subsection{Limitations and Deployment Scope}
\label{sec:limitations}

The current construction has three limitations.

\noindent\textbf{Trust boundary}. The malicious protocol protects every corruption set in $\mathsf{Adm}_t$ under the designated sender-receiver non-collusion assumption also used by Gao et al.~\cite{glw24}. In software-supply-chain exposure coordination, an industry vulnerability coordinator or software supplier can operate $\mathcal{P}_1$, a national computer emergency response team or supervisory authority can operate $\mathcal{P}_2$, and participating operators can assume the remaining roles. Separate administration, legal duties, access control, and auditing support this boundary. The security theorem does not cover deployments in which the two designated roles may collude.

\noindent\textbf{Performance scope}. The evaluation in Tables~\ref{exp1} and~\ref{exp2} and Figure~\ref{exp3} shows that the largest gains occur for large inputs and large values of the corruption bound $t$, particularly in the malicious setting. For small inputs, the protocol of Nevo et al.~\cite{nty21} can have lower malicious runtime, Guan et al.~\cite{gga25} can be faster in the semi-honest model, and Gao et al.~\cite{glw24} give lower semi-honest cost in most evaluated settings. Our main performance advantage is therefore concentrated in deployments requiring malicious security across admissible roles, large-set processing, or high corruption tolerance.

\noindent\textbf{IGOPPRF extensions}. The core malicious MPSI protocol in Section~\ref{sec:mpsi} does not use cuckoo hashing. The MPSI-cardinality, threshold MPSI, and MPSI-sum protocols in Section~\ref{extensionmpsi} use cuckoo hashing~\cite{cuckoo} and currently have semi-honest security proofs. Pinkas et al.~\cite{prty20} show that revealing a cuckoo location can leak information about the receiver's other inputs, so the standard semi-honest transformation does not directly provide malicious security. A malicious extension requires a maliciously simulatable hashing layer and consistency between the hashed representations and the inputs supplied to IGOPPRF and 2PC.

\section{Conclusion}

This work introduces IGOPRF, a new primitive that gives a unified paradigm for MPSI in both the semi-honest and malicious models. Under the designated sender-receiver non-collusion assumption, our protocols tolerate arbitrary admissible collusions and naturally extend to several MPSI variants. This makes IGOPRF well suited to privacy-preserving data circulation scenarios, where multiple data holders need to jointly exploit their data without exposing the underlying raw datasets. Evaluations across network conditions and security settings show practical online performance, clear advantages in malicious and large-input settings, and stronger security coverage under admissible collusions.

\section*{Acknowledgments}

This work was supported by the National Key Research and Development Program of China (No. 2023YFB3106500), the National Natural Science Foundation of China under Grant (No. 62672305), the Program of China Scholarship Council (No. 202506230132), the Singapore Ministry of Education Academic Research Fund Tier 2 Grant (No. MOE-T2EP20223-0016), and the Key Project of Science and Technology Research by the Chongqing Education Commission (No. KJZD-K202400610).

\section*{Ethical Considerations}

This study develops and evaluates cryptographic protocols using synthetic inputs. It involves no human subjects, real user data, or live infrastructure. Deployments should follow applicable legal and organizational requirements and enforce access control, auditing, purpose limitation, and data minimization.

\bibliographystyle{IEEEtran}
\bibliography{paper}

\end{document}